\documentclass[fleqn,usenatbib]{mnras}
\usepackage{graphicx}	
\usepackage{amsmath}	
\usepackage{amssymb}	
\usepackage{comment}
\usepackage{xspace}
\usepackage{lastpage}

\def\Sref#1{$\S$\ref{#1}\xspace}

\def\Fref#1{Fig.~\ref{#1}\xspace}

\def\Tref#1{Table~\ref{#1}\xspace}

\def\Cref#1{Chapter~\ref{#1}\xspace}

\def\redmapper{\textsc{redMaPPer}\xspace}

\defcitealias{Diemer2014}{DK14}
\defcitealias{More2016}{M16}
\defcitealias{Yang2007}{Y07}

\title[The Boundaries of Halos in SDSS]{The Halo Boundary of Galaxy Clusters in the SDSS}

\author[Baxter, Chang et al.]{
Eric Baxter$^{1}$\thanks{E-mail: ebax@sas.upenn.edu},
Chihway Chang$^{2}$, 
Bhuvnesh Jain$^{1}$, 
Susmita Adhikari$^{3}$,  \newauthor
Neal Dalal$^{3,4}$,
Andrey Kravtsov$^{2,5,6}$, 
Surhud More$^{7}$,
Eduardo Rozo$^{8}$,  \newauthor
Eli Rykoff$^{9,10}$,
Ravi K. Sheth$^{1,11}$
\\ \\
$^{1}$Center for Particle Cosmology, Department of Physics, University of Pennsylvania, Philadelphia, PA 19104, USA\\
$^{2}$Kavli Institute for Cosmological Physics, The University of Chicago, Chicago, IL 60637, USA\\
$^{3}$Department of Astronomy, University of Illinois at Urbana-Champaign, Champaign, IL 61801, USA \\
$^{4}$Department of Physics, University of Illinois at Urbana-Champaign, Champaign, IL 61801, USA \\
$^{5}$Department of Astronomy and Astrophysics, The University of Chicago, Chicago, IL 60637, USA \\
$^{6}$Enrico Fermi Institute, The University of Chicago, Chicago, IL 60637, USA \\
$^{7}$Kavli Institute for the Physics and Mathematics of the Universe (WPI), Tokyo Institutes for Advanced Study, \\
The University of Tokyo, 5-1-5 Kashiwanoha, Kashiwa-shi, Chiba, 277-8583, Japan \\
$^{8}$Department of Physics, University of Arizona, Tucson, AZ 85721, USA\\
$^{9}$Kavli Institute for Particle Astrophysics \& Cosmology, P.O. Box 2450, Stanford University, Stanford, CA 94305, USA \\
$^{10}$SLAC National Accelerator Laboratory, Menlo Park, CA 94025, USA \\
$^{11}$The Abdus Salam International Center for Theoretical Physics, Strada Costiera 11, 34151 Trieste, Italy \\
}

\date{Last updated \today}
\pubyear{2017}

\begin{document}
\label{firstpage}
\pagerange{\pageref{firstpage}--\pageref{LastPage}}
\maketitle

\begin{abstract}
Mass around dark matter halos can be divided into ``infalling''
material and ``collapsed'' material that has passed through at least
one pericenter.  Analytical models and simulations predict a rapid
drop in the halo density profile associated with caustics in the
transition between these two regimes.  Using data from SDSS, we
explore the evidence for such a feature in the density profiles of
galaxy clusters and investigate the connection between this feature
and a possible phase space boundary.  We first estimate the steepening
of the outer galaxy density profile around clusters: the profiles show
an abrupt steepening, providing evidence for truncation of the halo
profile.  Next, we measure the galaxy density profile around clusters
using two sets of galaxies selected based on color.  We find evidence
of an abrupt change in the galaxy colors that coincides with the
location of the steepening of the density profile.  Since galaxies are
likely to be quenched of star formation and turn red inside of
clusters, this change in the galaxy color distribution can be
interpreted as the transition from an infalling regime to a collapsed
regime. We also measure this transition using a model comparison
approach which has been used recently in studies of the ``splashback''
phenomenon, but find that this approach is not a robust way to
quantify the significance of detecting a splashback-like feature.
Finally, we perform measurements using an independent cluster catalog
to test for potential systematic errors associated with cluster
selection. We identify several avenues for future work: improved
understanding of the small-scale galaxy profile, lensing measurements,
identification of proxies for the halo accretion rate, and other
tests.  With upcoming data from the DES, KiDS and HSC surveys, we can
expect significant improvements in the study of halo boundaries.
\end{abstract}

\begin{keywords}
galaxy: clusters: general -- cosmology: observations 
\end{keywords}

\section{Introduction}
\label{sec:intro}

In the standard cosmological model, gravitational collapse causes
small perturbations in an initially smooth dark matter density field
to collapse into dense clumps known as halos.  The matter distribution
in and around halos can be divided into two components, which we will
refer to as ``infalling'' and ``collapsed.''  Infalling material is in
the process of falling towards the halo, but has not yet passed
through an orbital pericenter.  Such material has experienced a {\it
  first turnaround} at the point when gravity halted its motion away
from the halo due to the expansion of the universe, but has not yet
experienced a {\it second turnaround} after passing by the halo.
Collapsed material, on the other hand, has experienced at least one
orbital pericenter passage and is in orbit around the
halo\footnote{Note that infalling material could include collapsed
  subhalos that are falling into a larger halo.} \citep{Gunn1972,
  Fillmore1984, Bertschinger1985}.  Close to the halo center collapsed
material dominates the mass distribution, while far away from the halo
center infalling material dominates.  The transition between these two
regimes happens near the halo virial radius; the scale of first
turnaround, on the other hand, is about five times larger.

Using N-body simulations, \citet[][hereafter
  \citetalias{Diemer2014}]{Diemer2014} determined that stacked dark
matter halo density profiles exhibit a sharp decline near the
transition from the infalling regime to the collapsed regime.
\citetalias{Diemer2014} associated this feature with the second
turnaround of dark matter particles, which results in a density
caustic in the accreted material.  A caustic here refers to a narrow,
localized region of enhanced density; just beyond the second
turnaround caustic, the density declines rapidly, producing the
feature observed by \citetalias{Diemer2014}.  Owing to the connection
between the observed steepening of the profile and second turnaround,
this feature has recently been termed the {\it splashback feature}.
Subsequently, \citet{Adhikari2014} developed a simple model for the
location of the feature and confirmed the results from
\citetalias{Diemer2014}.  The idea that second turnaround is
associated with a caustic in the density profile dates back to work by
\citet{Fillmore1984} and \citet{Bertschinger1985}.  However, it was
not obvious that a clear feature resulting from second turnaround
would persist in realistic simulations and after averaging across many
halos.

A significant steepening of the profile followed by a flattening as
one moves outward from the center is clearly seen in the clustering
signal of galaxies and clusters measured from Sloan Digital Sky Survey
(SDSS) data \citep{Abbas2007, Sheldon2009} and in weak lensing
measurements around SDSS clusters \citep{Johnston2007,Sheldon2009}.
Recently, \citet{Tully2015} presented evidence for a steep decline in
the galaxy density profile and a discontinuity in the velocity
dispersion around galaxy groups in a context similar to that
considered here.  Related investigations into cluster density profiles
in the infalling and collapsed regimes using spectroscopic data have
been performed by \citet{Rines2013}, \citet{Gifford2017} and
references therein.

\citet[][hereafter \citetalias{More2016}]{More2016} measured a
steepening of the galaxy density profile around \redmapper clusters
\citep{Rykoff2014, Rozo2014} identified in data from the SDSS eighth
data release (DR8) \citep{Aihara2011} and identified this steepening
with the splashback feature seen by \citetalias{Diemer2014}. By
fitting the \citetalias{Diemer2014} model for the radial density
profile --- which accounts for the rapid steepening of the profile
around the splashback radius --- to their SDSS measurements,
\citetalias{More2016} determined that the model with a splashback
feature provided a good fit to the data, while a model without a
splashback feature did not ($\chi^2$ of 60--140 for 9 degrees of
freedom).  \citetalias{More2016} then compared the location of the
feature inferred from the data to expectations from N-body
simulations, finding evidence of tension.

Identifying the splashback feature in data is challenging for many
reasons. First, observers typically measure only the projected density
profile of a halo rather than the 3D radial profile. Projection smears
out the otherwise sharp splashback feature, making it harder to
distinguish from a profile that does not have a splashback
feature. Second, while measuring the mass profile of halos is possible
with gravitational lensing, such measurements currently have
relatively low signal-to-noise. Measurements of galaxy density can be
used as a high signal-to-noise proxy for the matter density, but doing
so introduces additional uncertainties as the relation between the
galaxy density and the matter density is not known precisely. Third,
to increase the signal-to-noise of density profile measurements, one
typically stacks measurements across halos of a range of mass,
redshift and accretion rate. Stacking can broaden the sharp splashback
feature, making it more difficult to detect. Finally, effects such as
halo miscentering can introduce significant systematic uncertainties
into measurements of the halo profiles.

The main goal of this work is to carefully examine the transition from
the infalling to collapsed regimes around galaxy cluster halos using
data from SDSS.  In particular, we are interested in whether the data
provide evidence for a truncation of the halo density profile
consistent with that seen in simulations and whether such truncation
can be connected to the phase space behavior of the matter around the
halo.  These findings together would imply the existence a physical
halo boundary.  We employ the same SDSS-derived \redmapper cluster
catalog and galaxy catalog as used in \citetalias{More2016}. The large
number of \redmapper clusters and galaxies detected in SDSS make this
data set the best currently available for measuring the galaxy density
profile around clusters.  We extend the modeling of
\citetalias{More2016} to include an important source of systematic
error: the miscentering of halos in the cluster catalogs
\citep{George2012, vanUitert2015, Hoshino2015, Rykoff2014}.  Using
these improved models, we explore whether the data favor the truncated
Einasto model introduced by \citet{Diemer2014} to describe the
splashback feature over a pure Einasto model \citep{Sersic1963,
  Einasto1965}. We present constraints on the steepening of the
collapsed component of the halo profile near the splashback region and
compare with existing literature in both data and simulations.
Additionally, we investigate the relative abundance of red galaxies
around the same clusters as a signature of the transition from the
infalling to the collapsed regimes.  Finally, we perform similar
galaxy profile measurements using a cluster catalog derived from the
same SDSS data but independent of the \redmapper cluster catalog.
This test is important since it is conceivable that some feature of
the \redmapper algorithm could lead to the appearance of an artificial
splashback-like feature.  Concerns about potential systematic biases
affecting measurements of the splashback feature are well motivated:
recent work by \citet{Zu2016} suggests that the quantity $\langle
R_{\rm mem} \rangle$ used by \citetalias{More2016} to split their
cluster sample can be significantly contaminated by projection
effects.  A closer examination of the splashback feature in the
absence of $\langle R_{\rm mem} \rangle$ splitting is therefore
warranted.

As a brief aside, we note that the dark matter mass distribution is
commonly described using the halo model \citep[for a review
  see][]{Cooray2002}.  In the simplest version of this model, all of
the dark matter in the Universe is assumed to live inside of halos.
The matter distribution as measured by the halo-matter
cross-correlation can then be divided into two components: the
`one-halo' term, which describes the distribution of matter within
halos, and the `two-halo' term, which describes the distribution of
the halos themselves.  In the language of the halo model, the
``collapsed'' material can be associated with the one-halo term and
the ``infalling'' material can be associated with the two-halo term.
Halo models can also be written down in a way that make the connection
to phase space more explicit \citep[e.g.][]{Sheth2001}.  In this work,
however, we will generally use the terminology of the collapsed and
infalling material.

The paper is organized as follows. In \S\ref{sec:data} we summarize
the data used in this work; in \S\ref{sec:methods} we outline our
methodology for measuring and fitting the galaxy density profiles; in
\S\ref{sec:resultsi} we present our findings related to constraining
the halo profile in the infalling-to-collapsed transition regime; in
\S\ref{sec:resultsii} we present our results related to the relation
between this transition and the colors of galaxies; we discuss our
results and implications for future work in
\S\ref{sec:discussion}. Throughout this analysis we will assume a
flat-$\Lambda$CDM cosmological model with $h = 0.7$ and $\Omega_M
=0.3$ and we will measure cluster-centric distances in comoving units.
Logarithms in base 10 are denoted with $\log$.

\section{Data}
\label{sec:data}

We use data from SDSS in our analysis. The main dataset is the same as
that used by \citetalias{More2016}: the \redmapper galaxy clusters
described in \citet{Rykoff2014} and the SDSS DR8 photometric galaxies
\citep{Aihara2011}. We select galaxy clusters with richness
$20<\lambda<100$ and redshifts $0.1<z<0.33$, resulting in a catalog of
8649 clusters. The photometric galaxies are selected by requiring the
galaxy to have an $i$-band magnitude brighter than 21.0 (after dust
extinction correction), a magnitude error smaller than 0.1, and none
of the following flags: \textsc{saturated}, \textsc{satur\_center},
\textsc{bright}, and \textsc{deblended}. The Landy-Szalay estimator
used in \Sref{subsec:measurement} requires a set of random points that
uniformly populate the volume of space in which clusters and galaxies
could be observed. The cluster randoms used for this purpose are
generated by the \redmapper algorithm; these incorporate the redshift
and richness distribution of the \redmapper clusters. The galaxy
randoms were generated by distributing points uniformly inside the
footprint of the $i<21$ galaxy sample. 

To select the red and blue galaxies used in
\Sref{sec:resultsii}, we perform an additional color cut in
the rest-frame $g-r$ color.  We compute this quantity using the
$K$-corrected absolute magnitudes in the SDSS database,
\textsc{absMagG} and \textsc{absMagR}.  We define two subsamples: the
quartile with the largest $g-r$ (the ``red'' sample) and the quartile
with the smallest $g-r$ (the ``blue'' sample).  For the purposes of
this study, a more sophisticated selection based on e.g. a
red-sequence selection in color-magnitude-redshift space is not
necessary.  The simple selection defined here is sufficient to
demonstrate the connection between galaxy color and features of
interest in the galaxy density profile.

\section{Methods}
\label{sec:methods}

\subsection{Galaxy profile measurement}
\label{subsec:measurement}

We calculate the projected number density of galaxies around clusters,
$\Sigma_g(R)$, as a function of the projected comoving cluster-centric
distance, $R$, by cross-correlating the clusters and the galaxies.  We
compute the cluster-galaxy angular correlation using the Landy-Szalay
estimator \citep{Landy1993} in redshift bins of $\Delta z=0.05$. We
only bin the clusters in redshift bins; the galaxy photometric
redshift information is not used because of the large associated
uncertainties. For each redshift bin centered at $\bar{z}_{i}$, we
assume that clusters with $\bar{z}_{i}-\frac{\Delta
  z}{2}<z<\bar{z}_{i}+\frac{\Delta z}{2}$ are located at $z =
\bar{z}_{i}$. We then calculate the $i$-band absolute magnitude,
$M_{i}$, for all galaxies assuming that they are located at
$\bar{z}_{i}$. Following \citetalias{More2016}, we then restrict the
galaxy sample to $M_{i}-5\log(h)<-19.43$, corresponding to an apparent
magnitude cut of $m_i < 21$ at the redshift limit of the cluster
catalog, $z = 0.33$.  For each redshift bin, we measure the
cluster-galaxy correlation function in 15 comoving radial bins from
0.1 to 10.0 $h^{-1} {\rm Mpc}$.  The correlation function measurements
in a given redshift bin are then converted to $\Sigma_g$ by multiplying
by the mean galaxy density in that redshift bin.  Finally, we average
the measurements in all redshift bins, weighting by the number of
cluster-galaxy pairs in each bin. Similar to \citetalias{More2016}, we
use a jackknife resampling approach with 100 subregions to estimate
the covariance of our $\Sigma_g(R)$ measurement.

\citetalias{More2016} also measured the galaxy density profiles around
two subsamples of clusters split on the parameter $\langle R_{\rm
  mem}\rangle$, defined as the average of the cluster member distances
from the cluster center, weighted by the probability of cluster
membership.  $\langle R_{\rm mem} \rangle$ was first introduced by
\citet{Miyatake2016}, where it was shown that a sample of \redmapper
clusters split on this parameter exhibited similar masses (as inferred
from weak lensing observations), but different large scale clustering
biases, with the larger $\langle R_{\rm mem} \rangle$ sample having a
larger bias.  \citetalias{More2016} showed that the location of the
splashback radius inferred from their density profiles measurements
was correlated with $\langle R_{\rm mem} \rangle$.  Given the
connection between the splashback radius and cluster accretion rate
established by \citetalias{Diemer2014}, it was argued that $\langle
R_{\rm mem} \rangle$ could therefore provide a measure of the cluster
accretion rate.  However, recent work by \citet{Zu2016} suggests that
$\langle R_{\rm mem}\rangle$ is strongly affected by projection
effects which are in turn correlated with the surrounding density
field.  Given these concerns, we do not rely on $\langle R_{\rm
  mem}\rangle$ splits in this analysis.

\subsection{Modeling the splashback feature}
\label{subsec:model}

\citetalias{Diemer2014} measured the stacked density profile of dark
matter halos in simulations.  They fit an Einasto model
\citep{Einasto1965,Navarro2004} to the inner halo profile (radii $r
<0.5R_{\rm vir}$) while using the relation of \citet{Gao2008} to fix
the Einasto parameter $\alpha$ as a function of halo peak height,
$\nu$.  Extending these fits to the outer profile ($r > 0.5R_{\rm
  vir}$), \citetalias{Diemer2014} found that the stacked density
profiles exhibited a sharp decline relative to the Einasto fit just
outside the halo virial radius; this decline was associated with the
caustic produced by splashback of dark matter particles. To model this
behavior, \citetalias{Diemer2014} introduced simple fitting
formulae. They model the halo density profile as the sum of an Einasto
profile that effectively describes the collapsed material and a power
law profile that effectively describes the infalling
material\footnote{The \citetalias{Diemer2014} model also includes a
  constant term equal to the mean density of the Universe.  Here,
  since the measurements are effectively mean-subtracted, we do not
  include such a constant term.}.  The use of an Einasto profile to
model the collapsed material is well motivated by many studies using
N-body simulations \citep{Navarro2004, Merritt2005, Merritt2006,
  Navarro2010}.  The use of a power law term to describe the infalling
material is motivated by e.g. the self-similar collapse models of
\citet{Gunn1972}.  For a single peak, self-similar collapse models
predict a power law profile with index -1.5.  However, for CDM halos
forming as a result of gravitational collapse around intially Gaussian
perturbations, the infalling material is not expected to follow a pure
power law profile at large scales.  Furthermore, non-linear dynamics
can modify the profile of infalling material within the halo.  The
precise form of the infalling material profile must therefore be
calibrated using e.g. N-body simulations.  The simple power law model,
however, was shown to provide a good fit to the stacked profiles of
simulated halos out to $\sim 9 R_{\rm vir}$ in
\citetalias{Diemer2014}.  To model the observed steepening of the
density profile near $R_{\rm vir}$, \citetalias{Diemer2014} multiplied
the Einasto profile by the function $f_{\rm trans}(r)$, which is unity
for small $r$, but declines rapidly in a narrow region near the radius
$r_t$.

The complete profile introduced by \citet{Diemer2014} that provides
good fits to the stacked 3D density profile of simulated halos from
small scales out to $\sim 9 R_{\rm vir}$ has the form:
\begin{eqnarray}
\rho(r) &=& \rho^{\rm coll}(r) + \rho^{\rm infall}(r), \label{eq:rho} \\
\rho^{\rm coll}(r) &=& \rho^{\rm Ein}(r) f_{\rm trans}(r) \label{eq:einasto} \\
\rho^{\rm Ein}(r) &=& \rho_s \exp \left(-\frac{2}{\alpha}\left[\left( \frac{r}{r_{s}}\right)^{\alpha} -1\right] \right), \\
f_{\rm trans}(r) &=& \left[1 + \left(\frac{r}{r_t} \right)^{\beta} \right]^{-\gamma/\beta}, \label{eq:ftrans} \\
\rho^{\rm infall}(r) &=& \rho_0 \left(\frac{r}{r_0} \right)^{-s_e}, \label{eq:2halo}
\end{eqnarray}
where $\rho^{\rm coll}$ and $\rho^{\rm infall}$ represent the profiles of
the collapsed and infalling material, respectively.  Note that
$\rho^{\rm coll}$ and $\rho^{\rm infall}$ correspond to the $\rho_{\rm
  inner}$ and $\rho_{\rm outer}$ used by
\citetalias{Diemer2014}. Since $r_0$ is completely degenerate with
$\rho_0$, we will fix $r_0 = 1.5 \,h^{-1} {\rm Mpc}$ throughout.

The profile of Eqs.~\ref{eq:rho}--\ref{eq:2halo} contains eight free
parameters.  \citetalias{Diemer2014} first fit density profile
measurements from simulations allowing all eight parameters to vary
freely, and found that the profile provided a good fit to these
measurements.  Because some of the parameters in their fits were
correlated, \citetalias{Diemer2014} also explored how the number of
free parameters could be reduced by fixing various parameter
combinations.  In this analysis, we will allow all eight model
parameters (after fixing $r_0$) to vary independently for two reasons.
First, the parameter combinations constrained by
\citetalias{Diemer2014} depend on quantities such as the halo peak
height and the virial radius, both of which cannot be measured
precisely from the data.  Second, it is not necessarily true that
parameter combinations that can be fixed when fitting the dark matter
alone can also be fixed when fitting the galaxy distribution, given
the uncertain relation between galaxies and mass.  Allowing all eight
parameters to vary simultaneously was also the approach taken by
\citetalias{More2016}.  As we will discuss below, however, allowing
all eight parameters to vary freely (with some weak priors) can make
distinguishing between models that have a truncation caused by $f_{\rm
  trans}$ and models that have $f_{\rm trans} = 1$ difficult.

Another common parameterization for modeling the density profiles of
dark matter halos is the Navarro-Frenk-White (NFW) profile of
\citet{Navarro1996}.  The NFW profile is also known to be a good fit
to simulated dark matter halos, although it may not be as successful
as the Einasto model at capturing the behavior of the inner halo
profile \citep{Navarro2004, Merritt2005, Merritt2006, Navarro2010}.
Since we do not have a very strong theoretical prior to prefer the
Einasto profile over the NFW profile in this analysis of galaxy
density profiles, we will also consider the impact on our splashback
fits of replacing the Einasto profile with the generalized NFW model
(gNFW):
\begin{eqnarray}
\label{eq:gnfw}
\rho_{\rm gNFW}(r) &=& \frac{\rho_{i}}{\left( \frac{r}{r_s} \right)^{\alpha_{\rm gNFW}} \left( 1 + \frac{r}{r_s} \right)^{3-\alpha_{\rm gNFW}}},
\end{eqnarray}
where $\rho_i$ sets the normalization of the profile and $\alpha_{\rm
  gNFW}$ sets its shape.

Since we measure projected densities on the sky, it is necessary to
integrate $\rho(r)$ along the line of sight to obtain the projected
density $\Sigma(R)$:
\begin{eqnarray}
\Sigma(R) = \int_{-h_{\rm max}}^{h_{\rm max}} dh \, \rho(\sqrt{R^2 + h^2}),
\label{eq:Sigma_of_R}
\end{eqnarray}
where $R$ is the projected distance to the halo center. To avoid
divergence of the profiles, we restrict the line of sight integration
to $-h_{\rm max} < h < h_{\rm max}$.  We set $h_{\rm max} = 40\,h^{-1} {\rm Mpc}$, 
but find that our results are quite robust to this choice. 

The above equations for $\rho(r)$ and $\Sigma(R)$ were found to
accurately describe the mass distribution around simulated dark matter
halos in simulations by \citetalias{Diemer2014}. In this work,
however, we will follow \citetalias{More2016} and apply the same
models to the measured galaxy distributions, which we label with
subscript `g's: $\rho_g(r)$ and $\Sigma_g(R)$ (note that these
functions measure number densities rather than mass densities). That
is, we are assuming that any differences between the galaxy
distribution and the dark matter mass distribution (i.e. galaxy bias)
can be absorbed into the fitting parameters. In the limit of constant
galaxy bias, this assumption is certainly true.  However, at small
scales, galaxy bias is expected to be scale-dependent
\citep[e.g.][]{Seljak2000, Peacock2000} and as a result, this
assumption may break down. \citetalias{More2016} tested this
assumption using subhalo profiles around cluster-size halos in dark
matter simulations, showing that it is robust. However, the galaxy
density profile is not expected to follow the subhalo profile at small
scales, and the precise relation between the galaxy profile and the
matter profile on small scales is still an active research area
\citep[e.g.][]{Nagai2005, Guo2011, Budzynski2012}.

In the model testing parts of this work, we will adopt an operational
definition and define the splashback radius as the location of the
steepest slope in the model density profiles.  To differentiate
between the splashback radius in the 2D and 3D profiles, we define
$R_{\rm sp}^{3D}$ as the location of steepest slope in the
three-dimensional galaxy density ($\rho_g$), and $R_{\rm sp}^{2D}$
as the analogous quantity in the projected galaxy density
($\Sigma_g$). Note that alternate ways of identifying the splashback
radius exist in the literature \citep[e.g.][]{Mansfield2016}. Our
definition has the benefit of being well-defined and relatively easy
to measure in observational data.

\subsection{Modeling cluster miscentering}
\label{subsec:mis_center}

To measure the cluster-centric distance $R$ in the data we use the
cluster centers computed by the \redmapper algorithm. \redmapper
assigns cluster centers in a probabilistic fashion: each cluster
member galaxy is assigned a probability of being the cluster center,
$P_{\rm cen}$ based on its color, magnitude, redshift and local
density \citep{Rykoff2014}. The galaxy with the highest $P_{\rm cen}$
is then considered to be the cluster center. The model for the cluster
density profile introduced in \S\ref{subsec:model}, on the other hand,
is defined with respect to the center of dark matter halos identified
in N-body simulations using the \textsc{Rockstar} halo finder
\citep{Behroozi2013}.

It is possible for the \redmapper cluster center to differ from the
centers of \textsc{Rockstar}-identified halos in two ways. First, it
is possible that no cluster galaxy lies at the true center of the dark
matter halo. This can happen stochastically, or if observational
effects such as masking prevented the central galaxy from being
observed. A second possibility is that a cluster member galaxy does
lie at the true center of the dark matter halo, but it is not the
galaxy with the highest $P_{\rm cen}$. We refer to both of these
effects as {\it miscentering}. Miscentering can significantly alter
the measured density profile at scales below the typical miscentering
distance (i.e. the distance between the assumed and true halo
centers). Although the transition between the infalling and collapsed
regimes occurs at scales greater than the miscentering distance, we
will see below that changes to the small-scale halo profile can
significantly alter how models fit the profile in the transition
region.  We note that \citetalias{More2016} tested for the effects of
miscentering on their determination of the splashback radius by
selecting clusters with high $P_{\rm cen}$ and repeating the density
profile measurements; however, they did not include a prescription for
miscentering in their model for the galaxy density.

We model the effects of miscentering following the approaches of
\citet{Melchior2016} and \citet{Simet2016}. The miscentered density
profile, $\Sigma_g$, can be related to the profile in the absence of
miscentering, $\Sigma_{g,0}$, via
\begin{eqnarray}
\Sigma_g = (1-f_{\rm mis})\Sigma_{g,0}(R) + f_{\rm mis} \Sigma_{g, {\rm mis}}(R),
\end{eqnarray}
where $f_{\rm mis}$ is the fraction of clusters that are miscentered,
and $\Sigma_{g, {\rm mis}}$ is the galaxy density profile for the
miscentered clusters. For clusters that are miscentered by $R_{\rm
  mis}$ from the true halo center, the corresponding density profile
is \citep{Yang2006, Johnston2007}
\begin{eqnarray}
&&\Sigma_{g,{\rm mis}} (R|R_{\rm mis}) = \nonumber\\
&&\int_0^{2 \pi} \frac{d\theta}{2\pi} \Sigma_{g,0}\left(\sqrt{R^2 + R_{\rm mis}^2  + 2RR_{\rm mis}\cos \theta}\right). \nonumber \\
\end{eqnarray}
The profile averaged across the distribution of $R_{\rm mis}$ values is then
\begin{eqnarray}
\Sigma_{g,{\rm mis}}(R) = \int dR_{\rm mis} P(R_{\rm mis}) \Sigma_{g,{\rm mis}}(R|R_{\rm mis}),
\end{eqnarray}
where $P(R_{\rm mis})$ is the probability that a cluster is
miscentered by a (comoving) distance $R_{\rm mis}$. Following
\citet{Simet2016}, we assume that $P(R_{\rm mis})$ results from a
miscentering distribution that is a 2D Gaussian on the sky. The 1D
probability distribution $P(R_{\rm mis})$ is then given by a Rayleigh
distribution:
\begin{eqnarray}
P(R_{\rm mis}) = \frac{R_{\rm mis}}{\sigma_R^2} \exp\left[ -\frac{R_{\rm mis}^2}{2\sigma_R^2} \right],
\end{eqnarray}
where $\sigma_R$ controls the width of the distribution. Following
\citet{Simet2016}, we set $\sigma_R = \tau R_{\lambda}$, where
$R_{\lambda} = (\lambda/100)^{0.2} \, h^{-1} {\rm Mpc}$ and we adopt
the mean value of $\bar{R}_{\lambda} = 0.98 \, h^{-1}{\rm Mpc}$ for
our sample.  The miscentering model is then completely specified by
the parameters $f_{\rm mis}$ and $\tau$. To determine how uncertainty
on miscentering propagates to uncertainty on evidence for splashback
we will consider several different (reasonable) priors on $f_{\rm
  mis}$ and $\tau$ below.

The weak lensing analysis of \redmapper clusters by \citet{Simet2016}
assumed Gaussian priors of $f_{\rm mis} = 0.2 \pm 0.07$ and $\tau =
0.4\pm 0.1$, derived using results from \citet{Rykoff2014}.
\citet{Rykoff2014} quantified the miscentering of the SDSS \redmapper
clusters based on 82 and 54 X-ray selected clusters in the XCS
\citep{Mehrtens2012} and ACCEPT \citep{Cavagnolo2009} data sets,
respectively.  Follow up studies from \citet{Hoshino2015} examined
data from stripe 82, finding that the visually-determined centroid of
the \redmapper clusters agree fairly well with the
\redmapper-determined centroid. Our fiducial analysis uses the
\citet{Simet2016} priors, but we will also consider variations on
these priors, including a model without miscentering and a model where
the widths of the priors on the miscentering parameters are
doubled. Existing data sets used to infer the amount of cluster
miscentering are quite limited and systematics in
identifying/selecting the X-ray cluster centers can introduce
additional uncertainties in the inferred miscentering priors
\citep{George2012}. 

\subsection{Model fitting}
\label{sec:fitting}

We fit the model described in \Sref{subsec:model} and
\Sref{subsec:mis_center} to the data using a Bayesian approach. We
define a Gaussian likelihood, $\mathcal{L}$, via
\begin{eqnarray}
\label{eq:likelihood}
\ln \mathcal{L}(\vec{d} | \vec{\theta}) = -\frac{1}{2}\left(\vec{d} - \vec{m}(\vec{\theta}) \right)^T \mathbf{C}^{-1}  \left( \vec{d} - \vec{m}(\vec{\theta})  \right),
\end{eqnarray}
where $\vec{d}$ is the data vector of $\Sigma_g$ measurements,
$\vec{m}(\vec{\theta})$ is the model for these measurements evaluated
at parameter values $\vec{\theta}$, and $\mathbf{C}$ is the covariance
matrix of the data. The free parameters of the model are $\rho_0$,
$\rho_s$, $r_t$, $r_s$, $\alpha$, $\beta$, $\gamma$, $s_e$, and the
miscentering parameters $\tau$ and $f_{\rm mis}$.

To compute the posteriors on the model parameters given the likelihood
of Eq.~\ref{eq:likelihood} we run a Markov Chain Monte Carlo (MCMC)
analysis using \texttt{emcee} \citep{FM2013}.  Following
\citetalias{More2016}, when computing the posterior we impose Gaussian
priors on several model parameters: $\log \alpha = \log 0.2 \pm 0.6$,
$\log \beta = \log 4.0 \pm 0.2$, $\log \gamma = \log 6.0 \pm
0.2$. Note that the prior on $\alpha$ adopted here and in
\citetalias{More2016} is quite weak. The central value of $\alpha=0.2$
is motivated from N-body simulations: given the weak lensing mass
estimates of the \redmapper clusters, simulations predict that dark
matter halos of the same mass should have $\alpha \sim 0.2$
\citep{Gao2008}. However, what we measure is the galaxy density
profile and not the dark matter profile, which is significantly less
constrained.  Previous measurements in SDSS and simulations by
\citet{Masjedi2006} suggest that the galaxy density profile may be
significantly steeper than the dark matter profile at scales smaller
than $\sim$0.1 $h^{-1}$Mpc.  Simulations also show that $\alpha$ is
dependent on the halo peak height, $\nu$, which leads to a dependence
on halo mass and redshift \citep{Gao2008}.  Finally,
\citet{Dutton2014} have shown that there is significant scatter in
$\alpha$ between different halos, with $\sigma_{\log \alpha} = 0.16 +
0.03z$.  Our prior on $\alpha$ is wide enough that it has little
effect on our parameter constraints.  One could imagine, however,
imposing a tighter prior on $\alpha$ motivated by simulations; as we
will discuss below, such a prior can significantly impact model fits
to the measured density profiles.  The central values of the priors on
$\beta$ and $\gamma$ are motivated by the analysis of
\citetalias{Diemer2014}. Finally, we restrict $r_s \in [0.1,
  5.0]\,h^{-1} {\rm Mpc}$, $r_t \in [0.1, 5.0]\,h^{-1} {\rm Mpc}$ and
$s_e \in [1.0, 10.0]$.  Given the MCMC parameter chains, we can
compute the location of the minimum of the logarithmic derivative of
the 3D density profile, $R_{\rm sp}^{3D}$.  Note that when computing
$R_{\rm sp}^{3D}$, we use the profile {\it without} the modifications
for miscentering since we are interested in the true splashback radius
of the halo.

Throughout this analysis, we set the upper limit of the scales we fit
to be $8.0\,h^{-1} {\rm Mpc}$, since the model introduced in
\S\ref{subsec:model} is not expected to be a good fit much beyond
$\sim9R_{\rm vir}$, where $R_{\rm vir}$ is the halo virial radius
\citepalias{Diemer2014}, and for the cluster sample considered here
$R_{\rm vir} \sim 1\,h^{-1} {\rm Mpc}$.  At small scales, systematics
in the galaxy density measurements are expected to become
significant. This is especially important in this analysis as cluster
fields are inherently crowded.  Issues such as detection
incompleteness, photometry inaccuracy and blending can be important at
these scales \citep{Melchior2015, Melchior2016}. In addition, the
relation between galaxy distribution and dark matter distribution
become more complicated at small scales, and the model described in
\Sref{subsec:model} may no longer be sufficient. In
\citetalias{More2016}, a minimum scale of 0.1 $h^{-1} {\rm Mpc}$ was
used.  In this analysis, we will consider two choices of the minimum
scale.

As splashback corresponds to a steepening of the outer halo density
profile, it is worthwhile to consider model-independent approaches to
measuring this steepening.  The logarithmic derivative of the density
profile can be constrained in model-independent ways (e.g. the
Savitzky-Golay approach taken by \citetalias{More2016} and
\citealt{Adhikari2016}).  However, a steep feature in the 3D profile
will appear significantly less steep in the 2D projected profile.
Furthermore, miscentering can change the shape of the splashback
feature in the 2D profile.  These effects present a challenge for
non-parametric methods, since such methods can only be applied to the
measured 2D profile and cannot be used to infer the 3D profile or the
profile in the absence of miscentering.  Additionally, near the
splashback radius the infalling matter and collapsed matter make
roughly equal contributions to the total density profile, making
inferences about the collapsed material alone difficult with
non-parametric methods.

\subsection{Model comparison via the Bayesian odds ratio}
\label{sec:splashback_preference}

We first consider a model comparison approach to determine whether the
data support the existence of a splashback feature.  In such an
approach, one must take care to define the models with and without the
feature.  For the model with a splashback feature, we adopt the model
of \citetalias{Diemer2014}.  Defining a model without a splashback
feature is more complicated.  Since $f_{\rm trans}$ was introduced by
\citetalias{Diemer2014} to describe this feature, it makes sense to
consider $f_{\rm trans} = 1$ as a splashback-free model as was done by
\citetalias{More2016}.  However, in the simulations of
\citetalias{Diemer2014}, the Einasto parameters were fixed by fitting
only the inner halo ($R < 0.5 R_{\rm vir}$).  We fit the Einasto
profile as part of the global fit using all radii, and in this case
the steep outer profile around the splashback radius may drive the
best fit values of $r_s$ and $\alpha$ to smaller and larger values,
respectively.  This can both compromise the quality of the fit at
small radii and allow the Einasto profile in the $f_{\rm trans} = 1$
model to become sufficiently steep at $R \sim R_{\rm sp}^{2D}$ to
describe steepening due to a splashback feature.  For now, we will
perform the model comparison between models with $f_{\rm trans}$ free
and $f_{\rm trans} =1$, but we will return to the subtleties of this
comparison in \S\ref{sec:halo_slope}.

To compare the $f_{\rm trans} = 1$ and $f_{\rm trans} = {\rm free}$
models, we use a Bayesian odds ratio.  The odds ratio is defined as the
ratio of the posteriors for two models, $M_1$ and $M_2$, given the
data $D$ and prior information $I$ \citep[see e.g. ][]{Ivezic2014}:
\begin{eqnarray}
O_{21} \equiv \frac{p(M_2 | D,I)}{p(M_1 | D,I)} = \frac{P(D|M_2,I) P(M_2|I)}{P(D|M_1,I) P(M_1|I)}.
\end{eqnarray}
Assuming we have no prior reason to prefer $M_1$ over $M_2$, the above
reduces to
\begin{eqnarray}
\label{eq:odds_ratio}
O_{21} &=& \frac{P(D|M_2,I) }{P(D|M_1,I)}  \\
\label{eq:odds_ratio_integrals}
&=& \frac{\int p(D|M,\vec{\theta}_2,I)p(\vec{\theta}_2 |M_2,I) d\vec{\theta}_2}{\int p(D|M,\vec{\theta}_1,I)p(\vec{\theta}_1|M_1,I) d\vec{\theta}_1},
\end{eqnarray}
where $\vec{\theta}$ represents the parameter spaces of the two
models.  The terms in the numerator and denominator of
Eq.~\ref{eq:odds_ratio} are sometimes referred to as the {\it
  evidence} for $M_1$ and $M_2$, respectively.  The evidence measures
the probability that the data would be observed if a particular model
is correct.  The integrals in Eq.~\ref{eq:odds_ratio_integrals} are
high-dimensional, so to evaluate them we use a method that relies on
the MCMC parameter chains for both models\footnote{see
  \url{http://www.astroml.org/book_figures/chapter5/fig_model_comparison_mcmc.html}
  for an example.}. To interpret the odds ratios that we compute we
use the Jeffreys' scale \citep{Jeffreys1963}. The Jeffreys' scale
identifies the different regimes of $\ln(O_{21})$ with `weak'
($0<\ln(O_{21})<1.16$), `definite' ($1.16<\ln(O_{21})<2.3$), `strong'
($2.3<\ln(O_{21})<4.6$) and `very strong' ($4.6<\ln(O_{21})$)
preference for one model over another.  In addition to the odds ratio
we also report the $\Delta \chi^2$ between the two models in order to
compare with \citetalias{More2016}.

\subsection{Test with alternative cluster catalog}
\label{subsec:yang}

The construction of a cluster catalog like \redmapper requires many
non-trivial choices that can potentially lead to poorly understood
selection effects that could alter the measured $\Sigma_g$. To make
any claim of a splashback detection more robust, it is therefore
important to test the measurements on alternative cluster catalogs.  A
large number of cluster-finding algorithms exist in the literature
\citep{vanBreukelen2009, Soares-Santos2011, Wen2012, Rykoff2014}.

To this end, we repeat the measurements of $\Sigma_g$ using the group
catalog of \citet[][hereafter \citetalias{Yang2007}]{Yang2007}.  The
\citetalias{Yang2007} catalog contains groups of galaxies identified
in SDSS DR7 data across a wide range of masses, including clusters as
massive as the \redmapper clusters.  Unlike the \redmapper algorithm,
the \citetalias{Yang2007} algorithm does not require grouping galaxies
in color space \citep[see][ for a detailed description of the
  algorithm]{Yang2005}.  Instead, the \citetalias{Yang2007} algorithm
uses spectroscopic redshift information to iteratively assign galaxies
to groups.  Briefly, this iterative assignment assumes that all groups
live in dark matter halos whose masses are estimated from an assumed
mass-to-light relationship.  Given the estimated halo mass, galaxy
assignment is performed assuming that the galaxy distribution is
described by an NFW profile.  The halo centers and the mass-to-light
ratio are then adjusted and group assignment is iterated until
convergence is achieved. The scatter in the halo masses, $M_h$,
assigned in this fashion is expected to be roughly twice as large as
the scatter in the mass estimates obtained from the \redmapper
mass-richness relation \footnote{The most massive groups in
  \citetalias{Yang2007} are estimated to have a mass scatter of
  $\sim0.2$ dex. Using the \citet{Simet2016} mass-richness relation,
  the \redmapper mass scatter is expected to be roughly $\sim0.1$ dex,
  or half that of the \citetalias{Yang2007} mass estimates}.

We select groups in the \citetalias{Yang2007} catalog with halo masses
in the range $10^{13.90}< M_{h} <10^{14.89}$ $h^{-1}M_{\odot}$.
Accounting for the different mass definitions used by
\citetalias{Yang2007}, this mass range corresponds roughly to the mass
of the $20<\lambda<100$ clusters in \redmapper, assuming the
mass-richness relation derived in \citet{Simet2016}.  We impose a
redshift selection of $0.1<z<0.2$, where the upper cutoff comes from
\citetalias{Yang2007} and the lower bound is to match the \redmapper
catalog. The redshift range is slightly different from the \redmapper
cluster sample, but we do not expect significant redshift-dependences
of the splashback feature resulting from this difference. This
selection yields a total of 3292 groups, which is roughly 2.5 times
fewer than in the \redmapper cluster sample. This is mainly due to the
different redshift ranges of the two samples.

We measure the $\Sigma_g$ profile around the \citetalias{Yang2007}
groups using the galaxy sample described in \Sref{sec:data} and fit
these measurements using the models described in \Sref{subsec:model}.
For the miscentering model, since $f_{\rm mis}$ and $\tau$ are not
well constrained for the \citetalias{Yang2007} catalog, we adopt the
loose miscentering priors of Model D in \Tref{tab:evidence}.

\begin{figure*}
\centering
\includegraphics[width=0.9\linewidth]{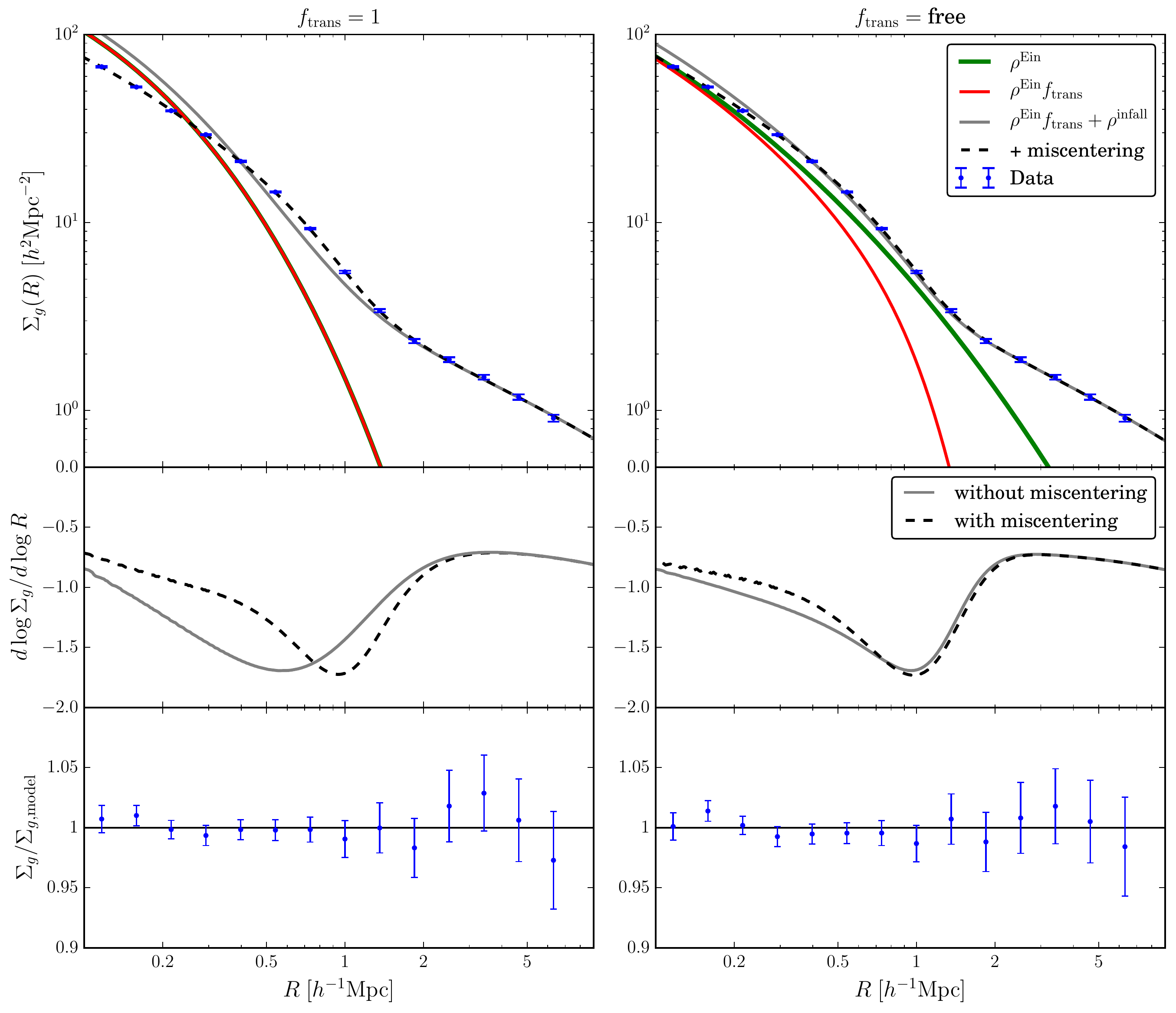}
\caption{The measured galaxy profile $\Sigma_g$ around \redmapper
  clusters in SDSS and the corresponding best-fitting models.  The top
  panels show $\Sigma_g$ measurements and model fits, the middle
  panels show the logarithmic derivative of the $\Sigma_g$ models, and
  the bottom panels show the ratio of the $\Sigma_g$ data points to
  the model.  The left panels shows the model fits with no steepening
  function (i.e. $f_{\rm trans}=1$), while the right panels show the
  fits with additional steepening beyond an Einasto profile
  (i.e. $f_{\rm trans}$ is allowed to vary).  The red curves in the
  upper panels show contribution to the projected galaxy density from
  the collapsed component ($\rho^{\rm coll}_g(r) = \rho^{\rm Ein}(r)
  f_{\rm trans}(r)$). The green curve in the right panel shows the
  contribution from the Einasto term of the model ($\rho^{\rm
    Ein}(r)$). The grey curves are the total profile without
  miscentering, while the dashed black curves are the profiles with
  miscentering.  Comparing the left and the right panels reveals that
  a model with large miscentering and $f_{\rm trans}=1$ can produce
  very similar total profile as a model with small miscentering and
  $f_{\rm trans}$ free.}
\label{fig:model_components_full}
\end{figure*}

\begin{table*}
\begin{center}
\caption{Results of model comparison with various modeling and data
  choices.  RM indicates the \redmapper catalog, \citetalias{Yang2007}
  indicates the catalog of \citetalias{Yang2007}.  $\Delta\chi^2$ and
  $\ln(O_{21})$ values indicate the results of the model comparison
  between the $f_{\rm trans} = 1$ and $f_{\rm trans} = {\rm free}$
  models, and are computed as described in
  \S\ref{sec:splashback_preference}.}
\begin{tabular}{|c|c|c|c|c|c|c|c|}
\hline
Model & Catalog & Priors &  Scales Fit [$h^{-1} {\rm Mpc}$] & $\Delta \chi^2$ & $\ln\left(O_{21}\right)$  \\ 
\hline 
A: no miscentering & RM & $f_{\rm mis} = 0.0$, $\tau = 0.0$  & $0.1 < R < 8.0$ & 139  & 69\\
B: fixed miscentering & RM & $f_{\rm mis} = 0.2$, $\tau = 0.4$ & $0.1 < R < 8.0$ &  73.3  & 36\\
C: miscentering with fiducial prior & RM & $f_{\rm mis} = 0.2 \pm 0.07$, $\tau = 0.4\pm 0.1$ & $0.1 < R < 8.0$ & 5.2  & 8.9\\
D: miscentering with wider prior & RM & $f_{\rm mis} = 0.2 \pm 0.14$, $\tau = 0.4 \pm 0.2$ & $0.1 < R < 8.0$ & 2.6 & 3.2\\
E: excluding small scales & RM &  $f_{\rm mis} = 0.2 \pm 0.07$, $\tau = 0.4 \pm 0.1$ & $0.3 < R < 8.0$ &  0.8  & 0.6 \\
F: NFW profile & RM &  $f_{\rm mis} = 0.2 \pm 0.07$, $\tau = 0.4 \pm 0.1$ & $0.1 < R < 8.0$ & 42.8  &  31 \\
\hline
\begin{tabular}
[x]{@{}c@{}}G: miscentering with wide priors,\\
tighter prior on $\alpha$
\end{tabular} 
& Y07 &  
\begin{tabular}
[x]{@{}c@{}} $f_{\rm mis} = 0.2 \pm 0.14$, $\tau = 0.4\pm 0.2$\\ 
$\log\alpha =\log(0.2) \pm 0.1 $ 
\end{tabular}  
& $0.1 < R < 8.0$ & 14.1  & 7.8 \\
\hline
\end{tabular}
\label{tab:evidence}
\end{center}
\end{table*}

\begin{table*}
\begin{center}
\caption{Best-fit model parameters with $f_{\rm trans}$ free (number
  preceding semicolon in each column) and $f_{\rm trans}=1$ (number
  following semicolon) and under different modeling assumptions.
  Modeling choices are described in \Tref{tab:evidence}. We have
  excluded some parameters in this table for clarity.  The remaining
  parameters are given in \Tref{tab:parameters_others}.}
\begin{tabular}{|c|c|c|c|c|c|c|c|c|c|}
\hline
Model & Catalog & $r_s [h^{-1} {\rm Mpc}]$ & $r_t [h^{-1} {\rm Mpc}]$ & $\alpha$ & $\beta$ & $\gamma$ & $f_{\rm mis}$ & $\tau$ & $R_{\rm sp}^{3D} [h^{-1} {\rm Mpc}]$ \\ \hline
A & RM & 0.85 ; 0.36 & 1.25 ; --- & 0.10 ; 0.42 & 3.83 ; --- & 6.26 ; --- & 0.0 ; 0.0 & --- ; --- & $1.23\pm0.05$\\
B & RM & 0.32 ; 0.29 & 1.31 ; --- & 0.16 ; 0.41 & 3.71 ; --- & 6.42 ; --- & 0.22 ; 0.22 & 0.32 ; 0.32 & $1.16\pm0.05$\\
C & RM & 0.27 ; 0.20 & 1.38 ; --- & 0.17 ; 0.41 & 3.98 ; --- & 6.73 ; --- & 0.22 ; 0.47 & 0.34 ; 0.40 & $1.18\pm0.08$\\
D & RM & 0.24 ; 0.19 & 1.42 ; --- & 0.19 ; 0.44 & 4.11 ; --- & 6.82 ; --- & 0.25 ; 0.51 & 0.34 ; 0.41 & $1.17\pm0.09$\\
E & RM & 0.35 ; 0.44 & 1.34 ; --- & 0.23 ; 0.93 & 3.66 ; --- & 6.45 ; --- & 0.20 ; 0.22 & 0.42 ; 0.43 & $1.15\pm0.07$\\
F & RM & 0.79 ; 0.10 & 1.23 ; --- & 1.54 ; 0.74 & 3.65 ; --- & 6.23 ; --- & 0.21 ; 0.50 & 0.45 ; 0.33 & $1.22\pm0.17$\\
\hline 
G &  Y07 & 0.35 ; 0.28 & 1.30 ; --- & 0.21 ; 0.38 & 3.75 ; --- & 6.20 ; --- & 0.51 ; 0.48 & 0.16 ; 0.20 & $1.16\pm0.08$\\
\hline
\end{tabular}
\label{tab:parameters}
\end{center}
\end{table*}

\section{Evidence for Halo Truncation}
\label{sec:resultsi}

\subsection{Model comparison tests and halo miscentering}
\label{sec:model_comparison_results}

The projected galaxy density profile, $\Sigma_g(R)$, measured around
\redmapper clusters is shown as the blue points with error bars in
\Fref{fig:model_components_full}.  Our measurement appears to be in
excellent agreement with that of \citetalias{More2016}.  We now
explore various model fits to these measurements and the results of
our model comparison tests.

We first consider as a check on our measurements the case where
miscentering is not included in the model and we fit the data over a
range of scales from 0.1 to 8.0 $h^{-1} {\rm Mpc}$. This is
essentially identical to the analysis of \citetalias{More2016}.  The
inferred $\Delta \chi^{2}$ and Bayesian odds ratio between the models
with $f_{\rm trans} = 1$ and $f_{\rm trans} = {\rm free}$ are listed
in \Tref{tab:evidence} (\redmapper Model A).  In this case, we find
significant preference for the model with $f_{\rm trans} = {\rm
  free}$, with $\Delta \chi^2 \sim 139$ relative to the $f_{\rm
  trans}=1$ model.  This value is consistent with the $\Delta \chi^2$
reported by \citetalias{More2016}. The evidence ratio also indicates
strong preference for the $f_{\rm trans} = {\rm free}$ model, having a
value of $\ln O_{21} \sim 76$, amounting to ``decisive'' evidence on
Jeffreys' scale.  The best-fit model parameters for the two models are
shown in \Tref{tab:parameters}.  Also listed in \Tref{tab:parameters}
are the constraints on $R_{\rm sp}^{3D}$ determined from the $f_{\rm
  trans} = {\rm free}$ model fits.  We note that our determination of
$R_{\rm sp}^{3D}$ is consistent with that of \citetalias{More2016}. We
do not list $R_{\rm sp}^{2D}$ here, but note that it is smaller than
$R_{\rm sp}^{3D}$ due to projection.

Next, we consider how miscentering affects the preference for the
$f_{\rm trans} = {\rm free}$ model.  We first explore the case where
the miscentering parameters $f_{\rm mis}$ and $\tau$ are fixed to the
central values from the priors of \citet{Simet2016} (Model B in
\Tref{tab:evidence}).  In this case, the $\Delta \chi^2$ and the
evidence ratio are decreased significantly relative to the
no-miscentering model, although both still indicate strong preference
for the $f_{\rm trans} = {\rm free}$ model. Comparing the Models A and
B for \redmapper in \Tref{tab:parameters}, we see that including
miscentering in the model fits changes the best-fit $r_s$ under the
model with $f_{\rm trans} = {\rm free}$ from 0.85 $h^{-1}\rm{Mpc}$ to
0.32 $h^{-1}\rm{Mpc}$.

When we allow the miscentering parameters to vary while imposing the
priors from \citet{Simet2016} (Model C), we find that the $\Delta
\chi^2$ and odds ratio are reduced even further. As shown in
\Tref{tab:parameters}, the $f_{\rm trans}=1$ model fits the data in
this case with a large miscentering fraction, $f_{\rm mis} = 0.47$.
This value is disfavored at roughly $3\sigma$ by the miscentering
priors of \citet{Simet2016}.  As we discuss below, we find that it is
precisely the miscentering prior that is largely driving the $f_{\rm
  trans} = 1$ model to be disfavored.

To help understand how miscentering is affecting the fits to
$\Sigma_g(R)$ and the results of the model comparison, we show the
model fits with the \citet{Simet2016} miscentering priors (Model C) in
\Fref{fig:model_components_full}.  The left panels of the figure
illustrate the fits with $f_{\rm trans }= 1$, while the right panels
illustrate the fits when $f_{\rm trans} = {\rm free}$ (note that in
the left panel, the green and red curves are identical since $f_{\rm
  trans}=1$).  Comparing the grey curves (which represent the total
model {\it without} miscentering) to the black dashed curves (which
represent the total model {\it with} the preferred miscentering), we
see that miscentering has the effect of flattening the inner galaxy
profile of the clusters. This makes sense: the offsets caused by
miscentering mean that the density profile is effectively averaged
across scales on the order of the miscentering radius, causing an
otherwise sharp inner profile to be flattened.  This redistribution of
the density at small scales also has the effect of {\it narrowing} the
minimum in the logarithmic derivative of the profile. 

\Fref{fig:model_components_full} makes it clear that the models with
$f_{\rm trans} = {\rm free}$ and with $f_{\rm trans}=1$ can both fit
the data very well (as further evidenced by the low value of $\Delta
\chi^2$ in this case).  The residuals for both fits (shown in the
bottom panels of the figure) appear almost identical between the two
model fits.  The preference for one model over the other, then, is
driven by the priors on the model parameters, in particular the prior
on $f_{\rm mis}$, as we will discuss below.  The minimum of the
logarithmic derivative of $\Sigma_g$ occurs in roughly the same
location in both fits, as can be seen in the middle panels of
\Fref{fig:model_components_full}. However, while the two models
generate very similar total profiles (and similar total logarithmic
derivatives), they fit the data in significantly different ways. To
see this, consider the red curves in \Fref{fig:model_components_full},
which show profile of the collapsed component when $f_{\rm trans}=1$
(left) and when $f_{\rm trans} = {\rm free}$ (right).  We see that the
$f_{\rm trans}=1$ model fits the outer profile ($R \gtrsim
0.5\,h^{-1}{\rm Mpc}$) with a large value of $\alpha$.  In general,
larger $\alpha$ results in a shallower inner profile for $r < r_s$.
However, in the case of the $f_{\rm trans} = 1$ fit, the value of
$r_s$ is decreased, which results in a steep profile at $R \lesssim
0.5\,h^{-1}{\rm Mpc}$; a large $f_{\rm mis}$ then flattens the inner
profile somewhat.  The model with $f_{\rm trans} = {\rm free}$, on the
other hand, prefers a shallower inner profile at the same radii (as a
result of larger $r_s$), does not require as much miscentering, and is
steepened substantially by the $f_{\rm trans}$ term at $R \gtrsim
0.5\,h^{-1}{\rm Mpc}$.

The model with $f_{\rm trans}=1$ prefers a miscentering fraction of
$f_{\rm mis} \sim 0.45$, in tension with the miscentering prior, which
prefers $f_{\rm mis} = 0.2$.  This tension between the preferred
miscentering fraction and the miscentering prior, when combined with
the behavior of the Einasto profile, drives the preference for the
model with $f_{\rm trans} = {\rm free}$ evaluated using the evidence
ratio.  In support of this conclusion, when we allow more freedom in
the miscentering model by doubling the widths of the \citet{Simet2016}
miscentering priors (Model D in \Tref{tab:evidence}), we find that the
evidence ratio in favor of $f_{\rm trans} = {\rm free}$ is weakened by
roughly a factor of 300 relative to the case with Model C miscentering
priors. Solely by going from a model without miscentering (Model A) to
a model with weak miscentering (Model D), the log odds ratio has been
reduced from $\ln O_{21} = 69$ to $\ln O_{21} = 3.2$.  We note,
though, that the odds ratio on the Jeffreys' scale is still
``strong'', even with these weaker miscentering priors.  When we
explore the extreme case of no priors on the miscentering parameters,
we find that the odds ratio is reduced to $\ln O_{21} = 0.86$,
amounting to only ``weak evidence.''

As noted previously, \citetalias{More2016} also considered the effects
of miscentering on their analysis, but took a very different approach
than that taken here.  \citetalias{More2016} repeated their
measurements of the galaxy density using only clusters of low
miscentering probability ($P_{\rm cen}>0.9$), finding that the change
in $R_{\rm sp}^{3D}$ was within measurement uncertainty. We have
repeated this test using our measurements, finding similar results.
However, we do find that the galaxy density profile for the high
$P_{\rm cen}$ clusters is somewhat steeper on small scales than for
the full sample, as is expected for a sample with better centering.
While miscentering may not significantly impact the location of
$R_{\rm sp}^{3D}$, as we have shown above, it can still have a
significant impact on the inferred model parameters and the shape of
the logarithmic derivative of the profile.  Furthermore, the $P_{\rm
  cen}$ parameter in \redmapper does not fully encapsulate all
possible mechanisms of cluster miscentering, such as the intrinsic
scatter between the center of the dark matter halo and the BCG (since
the \redmapper center is constrained to be on top of one of the
cluster galaxies).  For these reasons it is important to include
miscentering when modeling the halo profile as we have done here.

Given the impact of systematics such as miscentering, non-linear
galaxy bias, detection incompleteness, photometry inaccuracy and
blending on the inner density profile, it makes sense to consider
removing the innermost scales when fitting the galaxy density
measurements.  We perform such a fit by excluding scales below 0.3
$h^{-1} {\rm Mpc}$; the results are shown as Model E in
\Tref{tab:evidence} and \Tref{tab:parameters}.  We find that when
scales below 0.3 $h^{-1} {\rm Mpc}$ are excluded, the data no longer
exhibit a statistically significant preference for $f_{\rm trans} \neq
1$: $\Delta \chi^2 = 0.8$ and odds ratio of 1.8. This is not surprising 
given that this is effectively ignoring all the constraining power on small 
scales, which we have seen previously is very important for distinguishing 
the $f_{\rm trans}=1$ and  $f_{\rm trans}=$free models. 

\begin{figure}
\centering
\includegraphics[width=0.97\linewidth]{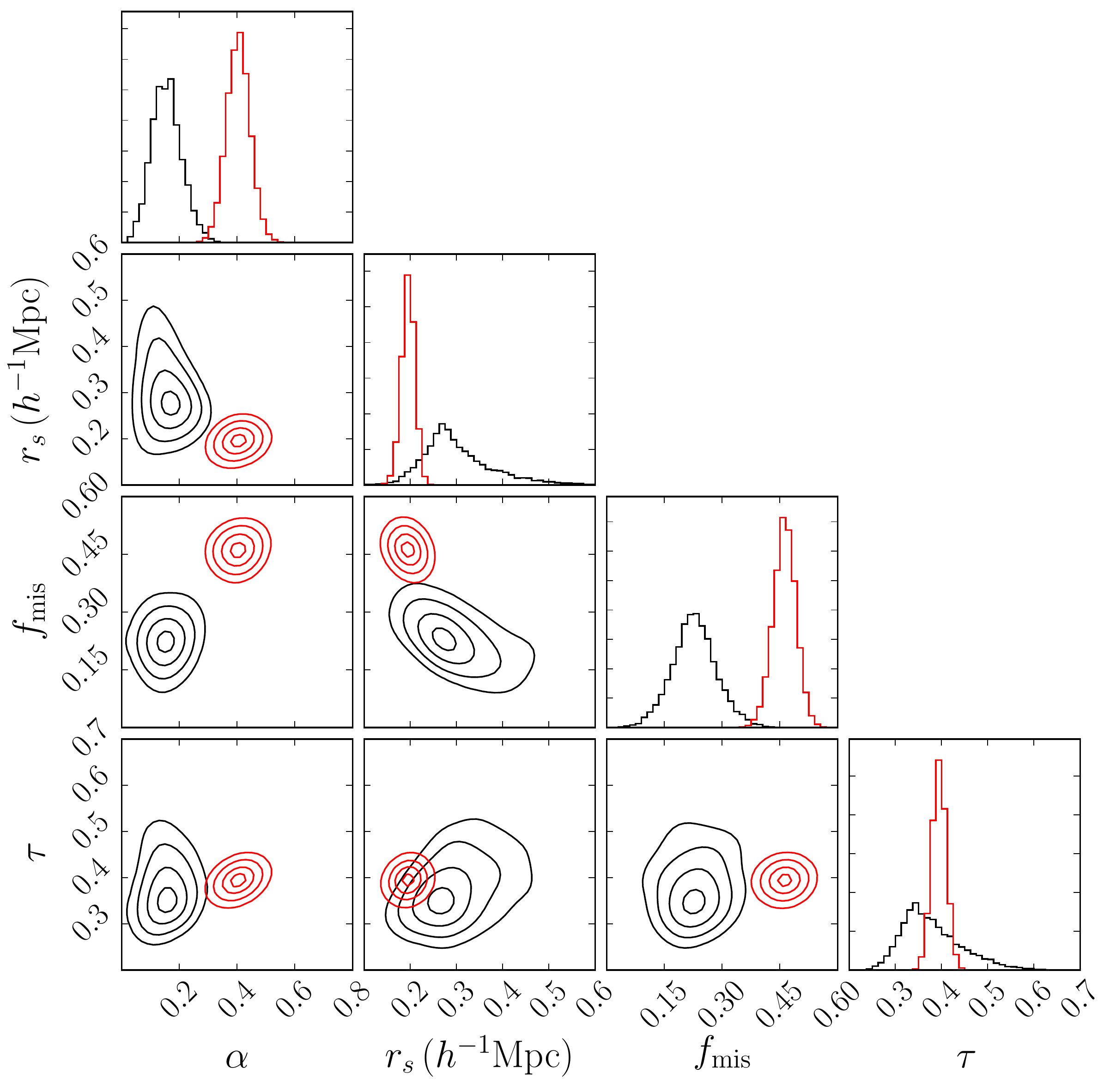}
\caption{Posteriors on the galaxy profile parameters recovered from
  the MCMC analysis of the galaxy profile measurements. Black curves
  show results of analysis that allows the parameters in the $f_{\rm
    trans}$ term of Eq.~\ref{eq:ftrans} to be free, while red curve
  shows results when $f_{\rm trans} = 1$.  Both analyses use the Model
  C miscentering priors from \Tref{tab:evidence}.}
\label{fig:posteriors}
\end{figure}

The above analysis highlights the fact that allowing additional
freedom in the inner galaxy density profile significantly affects the
ability of the data to distinguish between models with $f_{\rm trans}
=1$ and $f_{\rm trans} \neq 1$.  This behavior can be understood in
the following way.  The Einasto model of Eq.~\ref{eq:einasto} couples
the inner profile and the outer profile: as $\alpha$ is increased, the
inner profile becomes shallower while the outer profile becomes
steeper. If one ignores miscentering in the modeling of the galaxy
density profile (as was done in \citetalias{More2016}) then the value
of $\alpha$ is strongly constrained by the inner profile to be $\alpha
\sim 0.2$.  In this case, fitting the data at intermediate scales
requires truncation of the Einasto profile by the $f_{\rm trans}$
term, and $f_{\rm trans} = 1$ will be excluded at high significance.
If, on the other hand, one allows for miscentering (or removes the
innermost scales), the inner density profile can be fit by larger
$\alpha$, smaller $r_s$ and larger $f_{\rm mis}$.  Since a model with
larger $\alpha$ already has a steep outer profile, the preference for
additional steepening in the outer halo profile (as parameterized with
$f_{\rm trans}$) is reduced.  We compare the posteriors on $\alpha$,
$r_s$ and $f_{\rm mis}$ for the two model fits in
\Fref{fig:posteriors}.

We note that the high values of $\alpha$ preferred by the model fits
with $f_{\rm trans} = 1$ are disfavored by other studies. As shown in
\Tref{tab:parameters}, these fits generally prefer $\alpha$ in the
range 0.3--0.4.  These values are well within our prior of $\log
\alpha = \log 0.2 \pm 0.6$, but may be somewhat extreme relative to
expectations for the dark matter from N-body simulations
\citep{Gao2008}.  A tighter prior on $\alpha$ from a combination of
simulations and data would mean less sensitivity to the uncertainties
in the miscentering parameters, and would therefore help improve our
ability to make a more robust case for $f_{\rm trans} \neq 1$ using
the model comparison approach explored in this section.

Unlike the Einasto profile, the gNFW profile of Eq.~\ref{eq:gnfw}
forces the slope of the outer halo profile to asymptote to $-3$ at large
radius, regardless of parameter values.  Therefore, we expect that if
the measured outer halo profile exhibits a sharp steepening, the
preference for $f_{\rm trans} \neq 1$ will be greater when assuming an
gNFW profile than when using the Einasto model.  Indeed, as shown in
\Tref{tab:evidence} (Model F), the evidence for $f_{\rm trans} = 1$ is
increased when using the gNFW model.  In some sense, the gNFW analysis
provides a better measure of the detection significance of $f_{\rm
  trans} \neq 1$ because the outer slope is essentially fixed.
However, the sensitivity of the model comparison results to the
parameterization of the profile of the collapsed component is certainly
a drawback to this approach to detecting a splashback-like feature.

\begin{figure}
\centering
\includegraphics[width=0.927\linewidth]{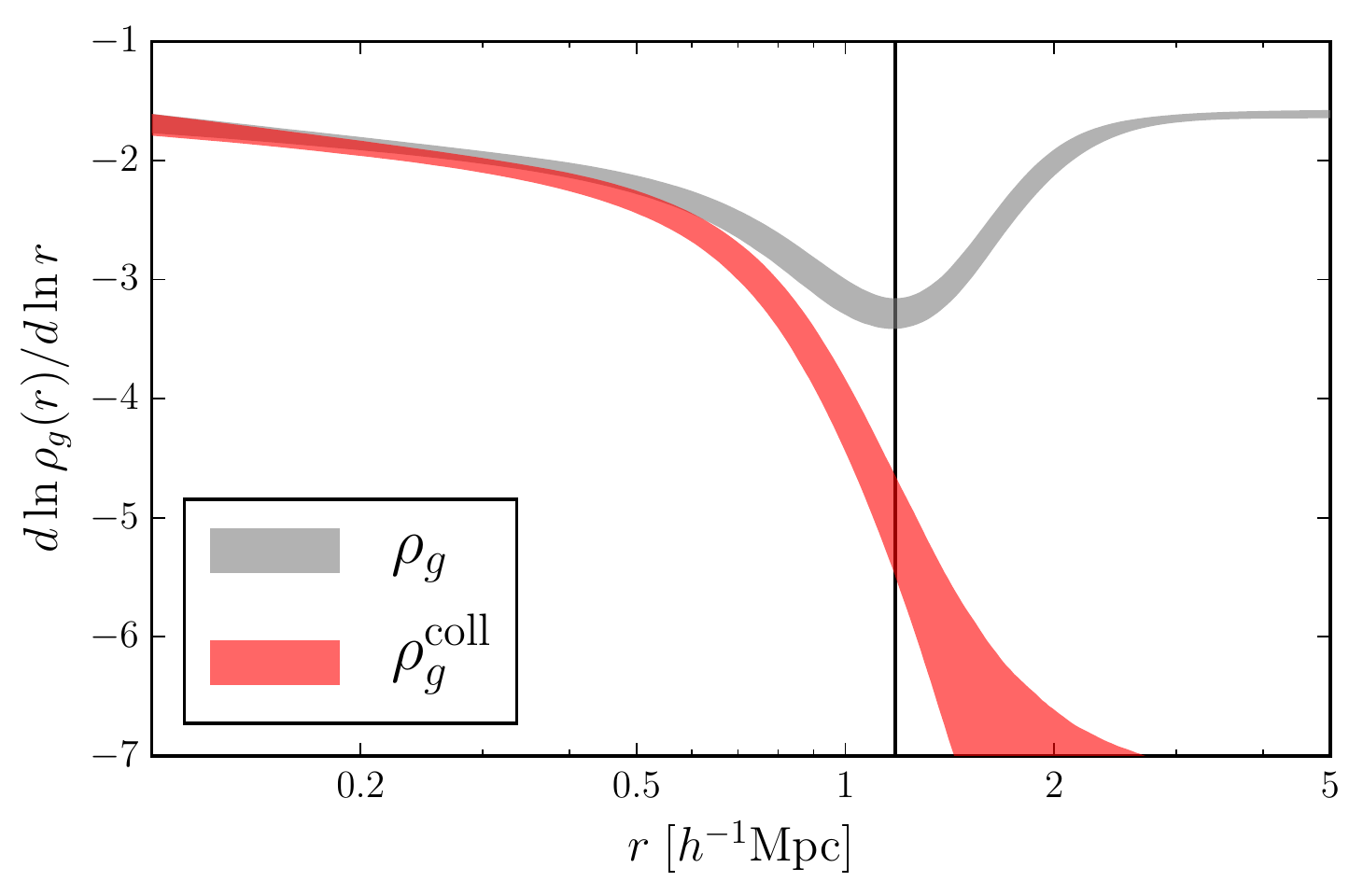}
\caption{Constraints on the 3D logarithmic derivative of the collapsed
  component ($\rho_g^{\rm coll}(r)$) and total galaxy density
  ($\rho_g(r)$) from our model fits to the measured galaxy density
  profile around \redmapper clusters.  The best fit value of the
  splashback radius, $R_{\rm sp}^{3D}$, is shown as the vertical line.
  The data prefer a profile which exhibits a steepening to slopes
  significantly steeper than $-3$ over a narrow range in radius.  This
  finding can be interpreted as evidence for truncation of the halo
  profile consistent with that seen in simulations by
  \citetalias{Diemer2014}.}\label{fig:logderiv_onehalo}
\end{figure}

\begin{figure}
\centering
\includegraphics[width=0.85\linewidth]{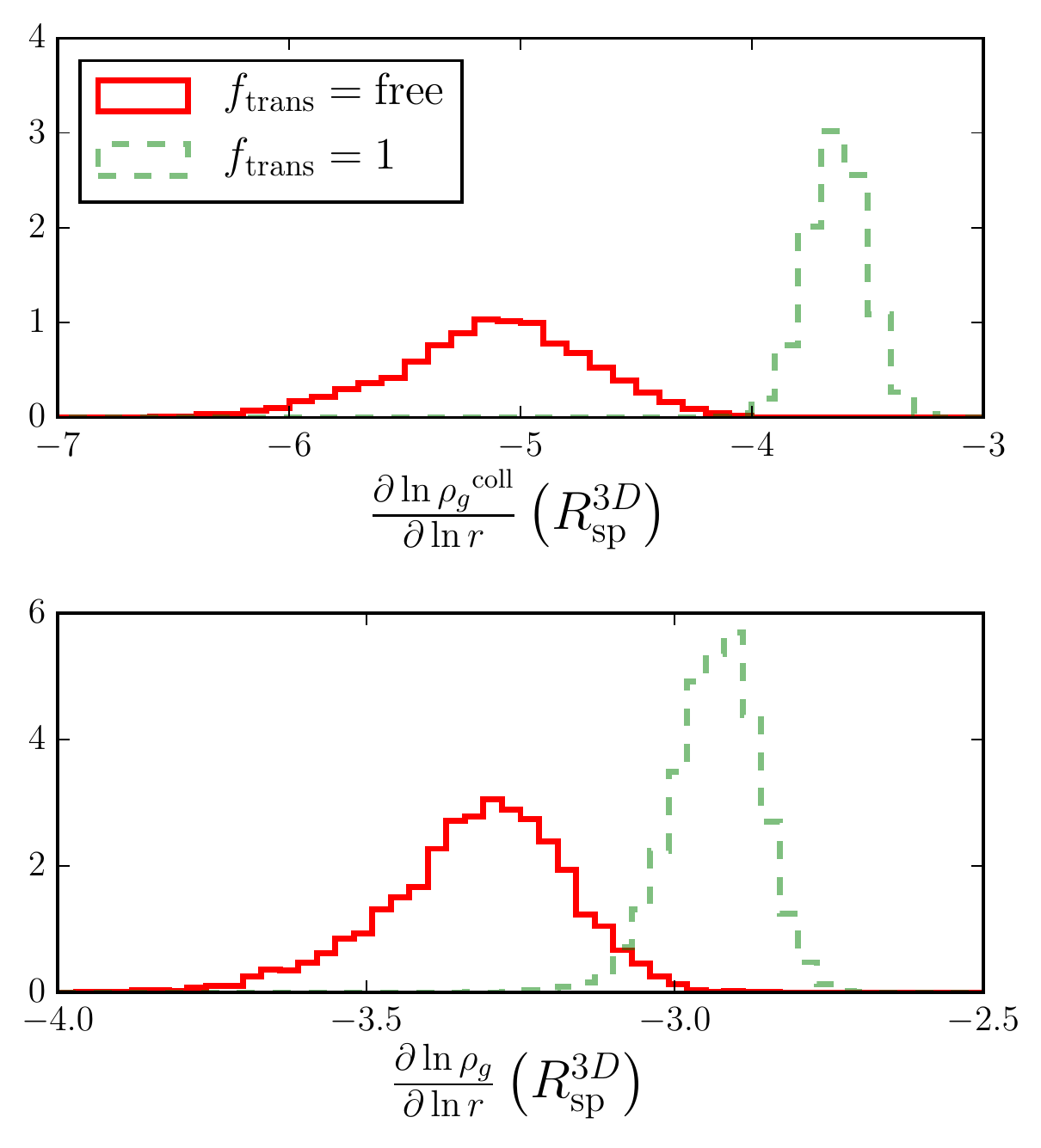}
\caption{Constraint on the logarithmic derivative of the three
  dimensional profile of collapsed component (top panel) and the total
  galaxy profile (bottom panel) evaluated at the splashback radius,
  $R_{\rm sp}^{3D}$, inferred from model fits to the measured galaxy
  density profiles. Solid (red) curves show results for fits when the
  parameters describing the profile of the splashback feature
  (i.e. $f_{\rm trans}$) are allowed to vary, while dashed (green)
  curves show results for fits with $f_{\rm trans} = 1$.  As discussed
  in the text, the $f_{\rm trans} = {\rm free}$ model provides a
  better description of the data.  We find significant evidence for
  slopes of the collapsed material profile significantly steeper than
  $-3$ at the splashback radius, suggesting a truncation of the halo
  profile. }
\label{fig:slope_dist}
\end{figure}

\subsection{The slope of the halo profile near $R_{\rm sp}$}
\label{sec:halo_slope}

We have seen that once miscentering is introduced into the $f_{\rm
  trans} = 1$ model from \citetalias{Diemer2014}, the resultant model
is flexible enough to produce a steep outer profile while still being
consistent (at moderate significance) with our fiducial priors.
Consequently, the ability of the data to distinguish between this
model and the $f_{\rm trans} = {\rm free}$ model is reduced.  An
alternative to the model comparison approach is to use the model fits
to directly constrain the logarithmic derivative of the halo profile
in the infalling-to-collapsed transition regime, as shown in
\Fref{fig:logderiv_onehalo}.  Using simple collapse models,
\citet{Dalal2010} argued the power law index of the outer halo profile
is set by the profile of the initial perturbation that gave rise to
the halo.  For power law initial profiles of arbitrary steepness, the
steepest logarithmic slope of the collapsed outer halo profile is
$-3$.  Departures from the power law form can give rise to logarithmic
slopes slightly steeper than $-3$, and for an NFW profile the slope
asymptotes to $-3$ at large radii.  Finding that the halo profile
reaches logarithmic slopes significantly steeper than $-3$ over a
narrow range of radii, then, would imply truncation of the halo
profile.\footnote{Halos that are simulated into the future and left to
  relax exhibit truncated Hernquist profiles \citep{Hernquist1990}
  that reach slopes of $-4$, however this steepening happens gradually
  rather than over a narrow region around $R_{\rm sp}$
  \citep{Busha2005}.}

To constrain the logarithmic slope of the galaxy density profile, we
draw sample profiles from our MCMC chains for the fits with $f_{\rm
  trans} = {\rm free}$ and compute the logarithmic derivatives of
these profiles.  \Fref{fig:logderiv_onehalo} shows the resultant
constraints on the logarithmic derivatives of $\rho_g(r)$ (grey band)
and $\rho_g^{\rm coll}(r)$ (red band).  The figure shows that we obtain a
fairly tight constraint on the slope of $\rho_g^{\rm coll}$ out to radii
at least as large as $R_{\rm sp}^{3D}$ (i.e. where the logarithmic
derivative of the total profile has a minimum, marked with a vertical
line in the figure).  This is a non-trivial finding since $\rho_g^{\rm
  coll}$ and $\rho_g^{\rm infall}$ have roughly equal magnitudes in this
regime and therefore make degenerate contributions to the total
profile.  It is clear from \Fref{fig:logderiv_onehalo} that over a
narrow range of radius (roughly $0.7\,h^{-1}{\rm Mpc}$ to
$1.3\,h^{-1}{\rm Mpc}$) the profile of the collapsed component exhibts
a rapid steepening from logarithmic slopes shallower than $-3$ to
logarithmic slopes steeper than -5.  This rapid steepening to slopes
less than $-3$ can be taken as evidence for truncation of the halo
profile.

The use of the model fits to infer the logarithmic derivative of the
galaxy density profile is well motivated.  We have shown that these
models provide excellent fits to the data (as illustrated by the
residuals in \Fref{fig:model_components_full}) and the
parameterization used here is known to be a good fit to the splashback
feature in simulations \citepalias{Diemer2014}.  Additionally,
\citetalias{More2016} has shown using simulations that this approach
can be used to accurately recover the 3D profile from the projected 2D
profile.  Finally, given the considerable freedom we have allowed in
our model fits, we expect our constraints on the logarithmic
derivatives to be quite robust to modeling uncertainty.

To explore the results shown in \Fref{fig:logderiv_onehalo} further,
we show the posterior on the logarithmic sopes of $\rho_g^{\rm coll}$
(top panel) and $\rho_g$ (bottom panel) evaluated at $R_{\rm sp}$ in
\Fref{fig:slope_dist}.  We show the posteriors for model fits with
$f_{\rm trans} = {\rm free}$ (red curves) and for $f_{\rm trans} =1$
(green curves).  Note, though, that since the $f_{\rm trans}=1$ model
is a special case of the $f_{\rm trans} = {\rm free}$ model and since
we have shown that the latter model generally provides a better fit to
the data, the $f_{\rm trans} = {\rm free}$ model should provide a more
accurate representation estimate of the profile slope than the $f_{\rm
  trans} =1$ model.

\Fref{fig:slope_dist} makes it clear that the data prefer a
logarithmic slope of $\rho_g^{\rm coll}$ at $R_{\rm sp}$ that is quite
steep, indeed significantly steeper than the $-3$ expected for a NFW
profile; the recovered estimate of the slope at $R_{\rm sp}$ is $-5.1
\pm 0.4$.  The estimated slope of the total profile is also
significantly steeper than $-3$, with a recovered value of
$-3.32\pm0.15$.  Therefore, even allowing for considerable uncertainty
in the profile of infalling matter, these results suggest that the
profile of collapsed component reaches slopes significantly steeper
than $-3$ near $R_{\rm sp}$.  Again, these findings can be taken as
evidence for truncation of the halo profile similar to that seen in
simulations by \citetalias{Diemer2014}.  Unlike the model comparison
results discussed in \S\ref{sec:model_comparison_results}, these
findings are quite robust to priors on the model parameters and
choices of fitting scales.

\Fref{fig:slope_dist} also shows the results obtained from the model
fits with $f_{\rm trans} = 1$ (green dashed curves).  These fits yield
significantly shallower slopes than the model fits where $f_{\rm
  trans}$ is allowed to vary.  This may be surprising given that the
logarithmic slopes of the projected profiles have almost identical
steepness (see \Fref{fig:model_components_full}).  The explanation is
that projection and miscentering act to smooth out a steep feature in
the 3D profile, making it appear significantly less steep in the 2D
profile.  Even these fits, however, prefer slopes of $\rho_g^{\rm coll}$
at $R^{3D}_{\rm sp}$ that are significantly steeper than -3.  Note,
though, that $R^{3D}_{\rm sp}$ inferred from the $f_{\rm trans} = 1$
fits tends to be smaller than $R^{3D}_{\rm sp}$ inferred from the
$f_{\rm trans} = {\rm free}$ fits (see
Fig.~\ref{fig:model_components_full}).  Consequently, the
distributions shown in \Fref{fig:slope_dist} are not coming from the
same physical radius.  Since the slope of the Einasto profile gets
more negative with increasing radius, evaluating the $f_{\rm trans} =
{\rm free}$ model slope at the $R^{3D}_{\rm sp}$ of the $f_{\rm trans}
= {\rm free}$ fits would make the value of the slope more negative.

\subsection{Alternative cluster catalog}

We now consider the results of analyzing the measurements of
$\Sigma_g$ around the groups identified in the \citetalias{Yang2007}
catalog.  As noted above, the use of an alternative cluster catalog
provides an important systematics test for the existence of a
splashback-like feature in the data.

Since the miscentering parameters appropriate for the
\citetalias{Yang2007} groups are not known precisely, we use the wide
miscentering priors of Model D in \Tref{tab:evidence}.  However, given
these wide priors and the lower signal-to-noise of the $\Sigma_g$
measurement for the \citetalias{Yang2007} groups, we find that for our
fiducial analysis the model fits with $f_{\rm trans} = {\rm free}$
prefer values of $r_t$ that are at the edges of the prior on this
parameter.  We find that this can be prevented by imposing a somewhat
tighter prior on $\alpha$: $\log \alpha = \log(0.2) \pm 0.1$.  This
prior is still fairly loose relative to expectations from simulations,
and is consistent with the values of $\alpha$ recovered from the
analysis of \redmapper clusters when $f_{\rm trans}$ is allowed to
vary.

The $\Sigma_g$ measurements around the \citetalias{Yang2007} groups
are shown in \Fref{fig:model_components_yang} and the results for the
model fits are summarized in \Tref{tab:evidence} and
\Tref{tab:parameters}.  In general, we find that the parameter values
recovered from the \citetalias{Yang2007} fits agree quite well with
those from the analysis of \redmapper clusters.  The recovered
splashback radius is also in good agreement.  The
\citetalias{Yang2007} analysis exhibits a large evidence ratio in
support of the model with $f_{\rm trans} = {\rm free}$, but this is
somewhat misleading (at least in comparison to the \redmapper results)
because of the prior we have imposed on $\alpha$ for this analysis.

\Fref{fig:slope_dist_yang} shows the distribution of profile slopes
for the \citetalias{Yang2007} groups, analogous to
\Fref{fig:slope_dist} for the \redmapper clusters.  The slope
distributions shown in \Fref{fig:slope_dist} are quite consistent with
those from the \redmapper measurement.  They indicate logarithmic
slopes of $\rho_g^{\rm coll}$ at $R_{\rm sp}$ of $-4.9 \pm 0.7$,
significantly steeper than the slope expected for an NFW profile and
consistent with expectations for a splashback feature.

We note that this test does not completely exclude the possibility of
systematic effects introduced by the cluster finder (\redmapper or
\citetalias{Yang2007}) which could bias the location of $R_{\rm sp}$
or the steepness of the collapsed component at $R_{\rm sp}$.
Nevertheless the fact that both \redmapper and \citetalias{Yang2007}
strongly prefer logarithmic slopes that are significantly steeper than
$-3$ at $R_{\rm sp}$ is fairly convincing evidence that the finding of
truncation of the halo profile is robust.
 
\section{Connecting the Halo Boundary to Galaxy Colors}
\label{sec:resultsii}

Our analysis has until now focused on examining the total galaxy
density profile near the transition between the infall regime and the
collapsed regime.  Another approach to probing this transition is to
examine galaxy colors.  The passage of a galaxy through a cluster is
expected to quench star formation in the galaxy.  This process can
happen through several channels: gravitational interactions with other
galaxies or the cluster potential itself \citep[e.g.][]{Moore1996},
stripping of the gas in the galaxy as a result of ram pressure from
cluster gas \citep[e.g.][]{Gunn1972}, and stripping of gas from the
galaxy's gaseous halo, thereby preventing replenishment of gas used to
form stars \citep[i.e. ``strangulation'', see e.g.][]{Larson1980,
  Kawata2008}.  Regardless of how it happens, quenching of star
formation will cause a galaxy to appear redder than galaxies with
active star formation.  Measurements of a transition in galaxy
clusters near the cluster virial radius have been performed in several
previous studies including \citet{Dressler1997}, \citet{Weinmann2006}
and references therein.  Here, we focus on the shape of this
transition and its connection to the phase space boundary between the
infalling and collapsed regimes.

\begin{figure}
\centering
\includegraphics[width=0.927\linewidth]{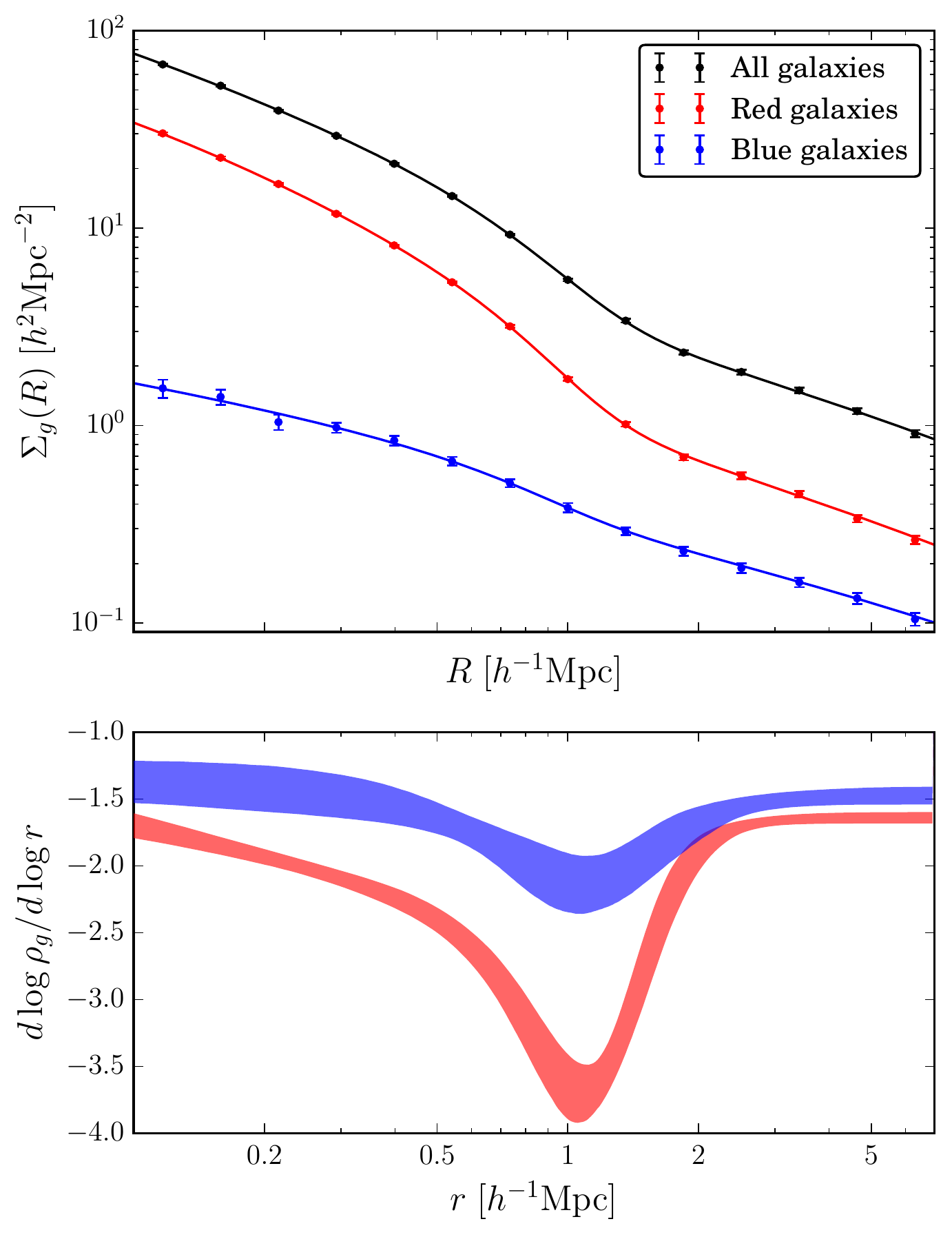}
\caption{The top panel shows the $\Sigma_g(R)$ measurements for the
  full sample (black data points), the reddest quartile of galaxies
  (red data points) and bluest quartile of galaxies (blue points);
  best fit models to the different measurements are shown as solid
  lines.  The bottom panel shows the corresponding log-derivatives of
  $\rho_g(r)$ inferred from model fiting.
}\label{fig:sigmag_red_fraction}
\end{figure}

In these scenarios, the typical time scales for quenching are
comparable to the time taken to move across the extent of the cluster,
roughly 2-4 Gyr \citep{Wetzel2013}.  Therefore, if interactions within
the cluster are responsible for quenching, a galaxy that has undergone
a single passage through a cluster will appear redder than a galaxy
that has not yet passed through the cluster.  Since galaxies outside
the splashback radius are significantly more likely to still be on
their first infall, we expect a sharp increase in the fraction of red
galaxies near the splashback radius.

Another possibility is that galaxy color simply correlates with
formation time or the time of accretion of the galaxy onto the cluster
and is not affected by processes inside the cluster
\citep[e.g.][]{Hearin2015}. In this case a sharp increase in the red
fraction at $R^{3D}_{\rm sp}$ would still be expected because within
this radius the infalling galaxies suddenly start to be mixed with
galaxies which were accreted ~2-4 Gyrs ago, which would also result in
a sharp change in the red fraction. The main point in the context of
this paper is that in both scenarios the sharp increase in the red
fraction is associated with the transition from the infalling to
collapsed regimes at the splashback radius.

\begin{figure}
\centering
\includegraphics[width=0.9\linewidth]{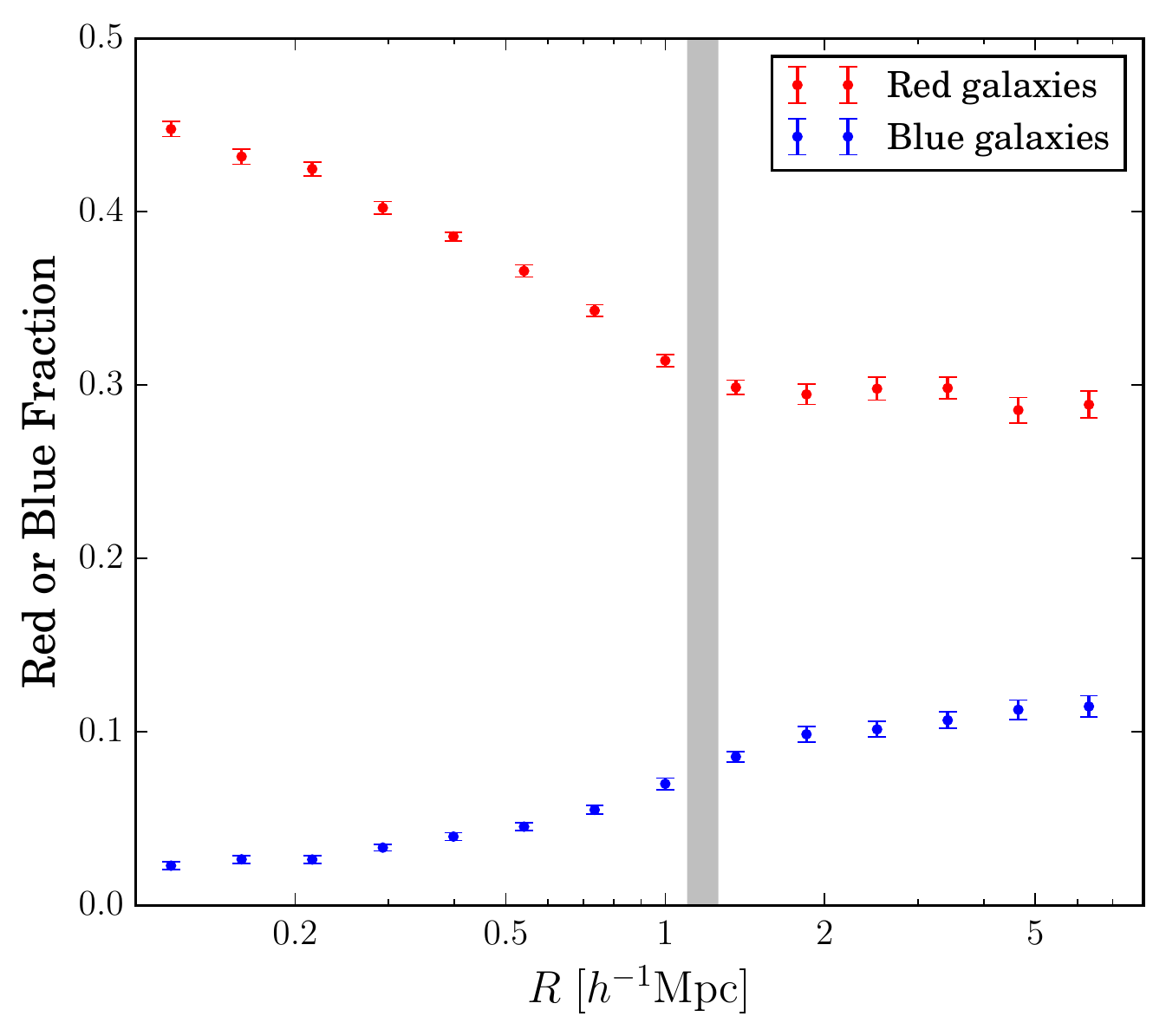}
\caption{The fraction of red and blue galaxies relative to all
  galaxies around \redmapper clusters as a function of the projected
  distance from the cluster center.
  }
\label{fig:red_fraction}
\end{figure}

To investigate this, we consider two galaxy subsamples selected based
on their rest-frame colors as described in \S\ref{sec:data}.  We
measure the galaxy density profiles of the individual subsamples, and
we will define the red/blue fraction to be the ratio of these two
profiles to the $\Sigma_g$ for the full galaxy sample.  The
$\Sigma_g$ and corresponding model fits for the two galaxy color
subsamples and the full sample are shown in the top panel of
\Fref{fig:sigmag_red_fraction}; the inferred logarithmic derivatives
of the 3D profile from the model fits are shown in the bottom panel.
For this analysis we use the Model C miscentering priors, as these
reflect recent constraints from analysis of \redmapper clusters.

Note that the profile of the blue galaxies approaches a power law with
index close to $-1.5$ at large scales.  Such a power law is precisely
the expectation for infalling material that has not reached shell
crossing \citep{Dalal2010}.  Consider particles outside of $R_{\rm
  sp}^{3D}$ that are falling towards a cluster which dominates the
local mass distribution.  In the absence of shell crossing, the mass
interior to the particles remains constant as they fall, and so their
free fall velocity scales as $v \sim r^{-0.5}$, where $r$ is the
distance between the cluster center and the particles.  The mass
contributed by such particles to a radial shell at $r$ and with
thickness $dr$ will be proportional to the time the particles spends
in the shell, so $\left(dM(r)/dr \right) dr \sim dr/v$, where $M(r)$
is the mass enclosed at radius $r$.  Assuming $M(r)$ follows a power
law and substituting the radial dependence of $v$, we have $M(r) \sim
r^{1.5}$.  The density profile, then, scales as $\rho \sim M/r^3 \sim
r^{-1.5}$.


The blue galaxy sample appears to be quite consistent with a purely
infalling component at large $r$.  In ``pre-processing'' models,
quenching occurs in the dense environment surrounding the cluster, but
prior to falling into the cluster \citep{Fujita2004}.  Quenching in
this way should remove galaxies from the blue sample as $r$ decreases
and add galaxies to the red sample, leading to a departure from the
expected slope of $-1.5$.  The fact that we observe logarithmic
derivatives close to -1.5 constrains the degree to which
pre-processing contributes to quenching.  Note, however, that we do
observe slightly steeper slopes for the red galaxies than for the blue
galaxies, which could be consistent with some amount of quenching due
to pre-processing.  This picture is complicated somewhat by the
presence of so-called ``backsplash'' galaxies that have passed through
the cluster (and may therefore have been quenched by processes inside
the cluster) but have been ejected as a result of gravitational
slingshot to several virial radii \citep{Wetzel2014}. We postpone more
in depth modeling of these scenarios to future work.

Next, we show the red/blue fraction measurements in
\Fref{fig:red_fraction}.  It is clear from \Fref{fig:red_fraction}
that the red fraction shows an abrupt steepening at around
$1.2\,h^{-1} {\rm Mpc}$.  In a scenario without a phase space boundary
between the infalling and collapsed regimes, it is hard to imagine how
a sharp upturn in the red fraction could arise at such large scales.
The gas densities at these radii are quite low; how would a galaxy
passing through the cluster outskirts know to become red at this
particular radius?  The picture of phase space caustics and quenching
by the cluster (or at some time after accretion onto the cluster)
provide a natural explanation for the observed red fraction behavior.
In this picture, the galaxy quenches after one or more passages
through the cluster, and the transition from outside $R_{\rm sp}$ to
inside $R_{\rm sp}$ marks the transition from a regime for which most
galaxies have never undergone a passage through the cluster to a
regime for which most galaxies have undergone passage through the
cluster.  In support of this picture, the location of the upturn in
the red fraction is in excellent agreement with the $R^{3D}_{\rm sp}$
inferred from the galaxy density measurements (shown as the grey band
in \Fref{fig:red_fraction}).  Note that the three dimensional
$R^{3D}_{\rm sp}$ is the relevant radius of comparison since it is
this radius that marks the physical phase space boundary in
projection.  The agreement between the projected red fraction
measurements and the 3D splashback radius is non-trivial.

\section{Discussion}
\label{sec:discussion}

\subsection{Evidence for a halo boundary}

Secondary infall models have predicted caustics in the phase space
distribution of particles being accreted onto halos since the early
studies of \citet{Fillmore1984} and \citet{Bertschinger1985}. It was
not clear however whether the disruptive processes in the formation of
cold dark matter halos would smear out caustic-like features,
especially when stacking across many halos.  Recently, using
simulations and analytic arguments, \citetalias{Diemer2014} and
\citet{Adhikari2014} identified a rapid steepening of the density
profile of stacked halos in N-body simulations that they associated
with a density caustic arising from the second turnaround of matter
particles, also known as splashback.  Recent work by
\citetalias{More2016} has presented evidence of a narrow steepening of
the galaxy density around \redmapper clusters detected in SDSS; such a
finding is consistent with expectations for a splashback feature.

In this work, we attempt to determine to what extent available data
support the existence of a halo boundary related to the presence of a
phase space boundary between infalling and collapsed contributions to
the total density profile.  Two analyses are presented here:
\begin{itemize}
\item 
We decompose the profile into ``infalling'' and ``collapased'' (or
1-halo) components and use a model fitting approach to estimate the
slope of the collapsed component near the transition between these two
regimes.  Near the location where the steepest slope of the total
profile occurs, the 1-halo profile reaches a logarithmic slope of
about -5 over a narrow range of radius.  This is significantly steeper
than the expectation of an NFW-like profile, and supports the idea of
a truncated halo profile.\\
\item 
The second evidence is that the location of steepest slope of the
total profile coincides with an abrupt increase in the fraction of red
galaxies.  Presumably a fraction of galaxies inside the halo are
quenched due to one or more passages through the cluster.  In the
infalling regime, on the other hand, galaxies have (for the most part)
never been inside the cluster and are therefore likely to have at most
a gradual trend in red fraction.  
\end{itemize}
The results of these two analyses, shown in
\Fref{fig:logderiv_onehalo} and \Fref{fig:red_fraction}, lend support
to the presence of a phase-space halo boundary associated with a sharp
decline in the halo density profile.  In contrast with common
definitions of halo boundaries, such as $R_{200}$ or $R_{\rm vir}$, a
phase-space halo boundary at $R_{\rm sp}$ would constitute a real,
physical boundary to the halo \citep{More2015}.

As a systematics test of these findings, we repeat the measurements of
galaxy density around groups identified in the \citetalias{Yang2007}
catalog.  We find that these measurements prefer similar slopes of the
halo profile in the infalling-to-collapsed transition region as the
\redmapper clusters, as shown in \Fref{fig:slope_dist_yang}.  Recent
work by \citet{Zu2016} found that the $\langle R_{\rm mem} \rangle$
quantity used by \citetalias{More2016} to split their cluster sample
can be severely impact by projections along the line of sight, which
are in turn related to the local environment of the cluster.  Since
\citetalias{More2016} found large changes in the inferred splashback
radius with $\langle R_{\rm mem} \rangle$, one may worry about the
effects of projections on the measurement of the splashback feature.
Our analysis does not preclude the possibility that $R_{\rm sp}$ is
affected by projection effects or by selection effects inherent to
\redmapper.  However, while projections might smooth out an otherwise
sharp splashback feature or change the inferred $R_{\rm sp}$, it seems
unlikely that they could be responsible for the artificial appearance
of a splashback-like steepening.  On the other hand, it is possible
that some feature of the \redmapper algorithm could cause clusters
that exhibit a sharp decline in their profiles to be preferentially
selected, thereby leading to the appearance of a splashback feature.
Our measurement of a rapid steepening around the \citetalias{Yang2007}
groups, however, disfavors the possibility of the observed steepening
being due solely to a \redmapper artifact.  Since every cluster finder
is in principle susceptible to spatially dependent selection effects,
such a test cannot be definitive. However, extending the measurements
performed here to cluster samples selected based on e.g. Sunyaev
Zel'dovich decrement would provide an independent check on the optical
cluster samples used in this work.

Using the parameterized models for the splashback feature introduced
by \citetalias{Diemer2014} we have also attempted to quantify whether
the data support a model that has such a feature over a model that
does not.  We perform a model comparison by computing a Bayesian odds
ratio between a model that includes the steepening caused by a
splashback feature (parameterized via $f_{\rm trans}$ in
Eq.~\ref{eq:ftrans}) and a pure Einasto profile that does not have
additional steepening.  A similar approach was taken by
\citetalias{More2016} to quantify the significance of their splashback
detection, although only $\chi^2$ was reported there.  We extend the
modeling efforts of \citetalias{More2016} to include miscentering, an
important systematic affecting the galaxy density profile.  We find,
however, that given reasonably weak priors on the parameters of the
pure Einasto model, this model can come close to matching the
steepening of the model with additional steepening caused by
splashback.  We therefore conclude that until better data are
available or tighter priors on the model parameters can be obtained,
such a test is not a particularly useful way to quantify evidence for
a splashback-like feature in the data.

While basic uncertainties remain in connecting observations of a
projected density profile (further modulated by the relation between
galaxies and mass) with a boundary in phase space, we believe the
findings summarized above are a promising indication of a dynamically
motivated halo boundary. They provide avenues for studying physical
processes such as dynamical friction \citep{Adhikari2016}, dark energy
\citep{Stark2016}, self-interacting dark matter and modifications of
gravity \citepalias{More2016}.

\subsection{Future prospects}

In this analysis we have used the galaxy density profile and the red 
fraction in an attempt to infer something about the phase space
behavior near the boundaries of cluster halos.  There may be alternate
routes to probing this boundary that go beyond these two observables.
For instance, proxies for halo accretion rate would be of significant
utility to studies of splashback since the accretion rate is expected
to correlate with the location of the splashback feature.  While the
analysis of \citetalias{More2016} attempted to use $\langle R_{\rm
  mem} \rangle$ as a proxy for accretion, subsequent work by
\citet{Zu2016} has shown that this proxy is contaminated by projection
effects.  Another avenue, namely the impact of dynamical friction on
splashback, has been explored by \citet{Adhikari2016} and would be
valuable to test with future datasets.  Our measurement of the red 
fraction provides qualitative evidence for a phase space boundary.
However, more definitive statements about how the red fraction relates
to splashback will likely require further effort with models and
simulations of galaxies in cluster environments. The present work
motivates exploration of these various avenues for detecting the
splashback feature.

Ongoing and future galaxy surveys such as the Hyper
SuprimeCam\footnote{\url{www.naoj.org/Projects/HSC}} (HSC), the Dark
Energy Survey\footnote{\url{www.darkenergysurvey.org}} (DES), the Kilo
Degree Survey\footnote{\url{kids.strw.leidenuniv.nl}} (KiDS), and the
Large Synoptic Survey Telescope\footnote{\url{www.lsst.org}} (LSST)
will provide much higher statistical power to constrain splahback
models than the SDSS data. Together with better understanding of
priors for both the model of the galaxy profile and systematic effects
such as miscentering, we should expect significant improvement in our
ability to characterize the splashback feature in the near future.
The use of weak lensing measurements to more directly measure the
cluster mass profile and study splashback \citep[e.g.][]{Umetsu2016}
may also be fruitful, particularly with upcoming weak lensing data
sets from DES, HSC and KiDS.

Finally, we note that this work has not addressed in any detail the
location of the splashback feature.  Although our model fits differ
from those of \citetalias{More2016}, we do not find any significant
difference in the recovered values of $R_{\rm sp}$ (although we have
not performed any splits on $\langle R_{\rm mem} \rangle$ in this
analysis).  More exploration into the location of $R_{\rm sp}$ and
comparison to simulations is a fruitful avenue for future work.

\section*{Acknowledgements}

We thank Kathleen Eckert, William Hartley, Rachel Mandelbaum, Philip
Mansfield, Carles S\'anchez, Risa Wechsler, Simon White, and Ying Zu
for fruitful discussions while preparing this manuscript.  CC and AK
were supported in part by the Kavli Institute for Cosmological Physics
at the University of Chicago through grant NSF PHY-1125897 and an
endowment from Kavli Foundation and its founder Fred Kavli.  EB and BJ
are partially supported by the US Department of Energy grant
DE-SC0007901.

Funding for the Sloan Digital Sky Survey IV has been provided by
the Alfred P. Sloan Foundation, the U.S. Department of Energy Office of
Science, and the Participating Institutions. SDSS-IV acknowledges
support and resources from the Center for High-Performance Computing at
the University of Utah. The SDSS web site is www.sdss.org.

SDSS-IV is managed by the Astrophysical Research Consortium for the
Participating Institutions of the SDSS Collaboration including the
Brazilian Participation Group, the Carnegie Institution for Science,
Carnegie Mellon University, the Chilean Participation Group, the
French Participation Group, Harvard-Smithsonian Center for
Astrophysics, Instituto de Astrof\'isica de Canarias, The Johns
Hopkins University, Kavli Institute for the Physics and Mathematics of
the Universe (IPMU) / University of Tokyo, Lawrence Berkeley National
Laboratory, Leibniz Institut f\"ur Astrophysik Potsdam (AIP),
Max-Planck-Institut f\"ur Astronomie (MPIA Heidelberg),
Max-Planck-Institut f\"ur Astrophysik (MPA Garching),
Max-Planck-Institut f\"ur Extraterrestrische Physik (MPE), National
Astronomical Observatories of China, New Mexico State University, New
York University, University of Notre Dame, Observat\'ario Nacional /
MCTI, The Ohio State University, Pennsylvania State University,
Shanghai Astronomical Observatory, United Kingdom Participation Group,
Universidad Nacional Aut\'onoma de M\'exico, University of Arizona,
University of Colorado Boulder, University of Oxford, University of
Portsmouth, University of Utah, University of Virginia, University of
Washington, University of Wisconsin, Vanderbilt University, and Yale
University.

\bibliographystyle{mnras} \bibliography{splash_sdss}

\appendix

\section{Model parameter constraints}
\label{sec:model_parameters_more}

\Tref{tab:parameters_others} shows the fit results for the remaining
parameters not shown in \Tref{tab:parameters}.  

\begin{table*}
\begin{center}
\caption{Additional best-fit model parameters with $f_{\rm trans}$ free (number
  preceding semicolon in each column) and $f_{\rm trans}=1$ (number
  following semicolon) and under different modeling assumptions.}
\begin{tabular}{|c|c|c|c|c|}
\hline
Model & Catalog & $\rho_s [h^3 {\rm Mpc}^{-3}]$ & $\rho_0 [h^3 {\rm Mpc}^{-3}]$ & $s_e$  \\ \hline
A & RM & 3.76 ; 22.58 & 0.43 ; 0.27 & 1.61 ; 1.34\\
B & RM & 27.08 ; 37.15 & 0.43 ; 0.31 & 1.61 ; 1.43\\
C & RM & 36.99 ; 103.92 & 0.43 ; 0.39 & 1.61 ; 1.55\\
D & RM & 48.98 ; 121.73 & 0.43 ; 0.41 & 1.61 ; 1.58\\
E & RM & 23.35 ; 14.86 & 0.43 ; 0.43 & 1.61 ; 1.60\\
F & RM & 11.72 ; 2043.12 & 0.43 ; 0.12 & 1.61 ; 1.05\\
\hline 
G & Y07 & 15.75 ; 29.53 & 0.32 ; 0.21 & 1.63 ; 1.41 \\
\hline
\end{tabular}
\label{tab:parameters_others}
\end{center}
\end{table*}

\section{Galaxy density measurement around Y07 groups}
\label{sec:yang_plot}

\begin{figure}
\centering
\includegraphics[width=0.9\linewidth]{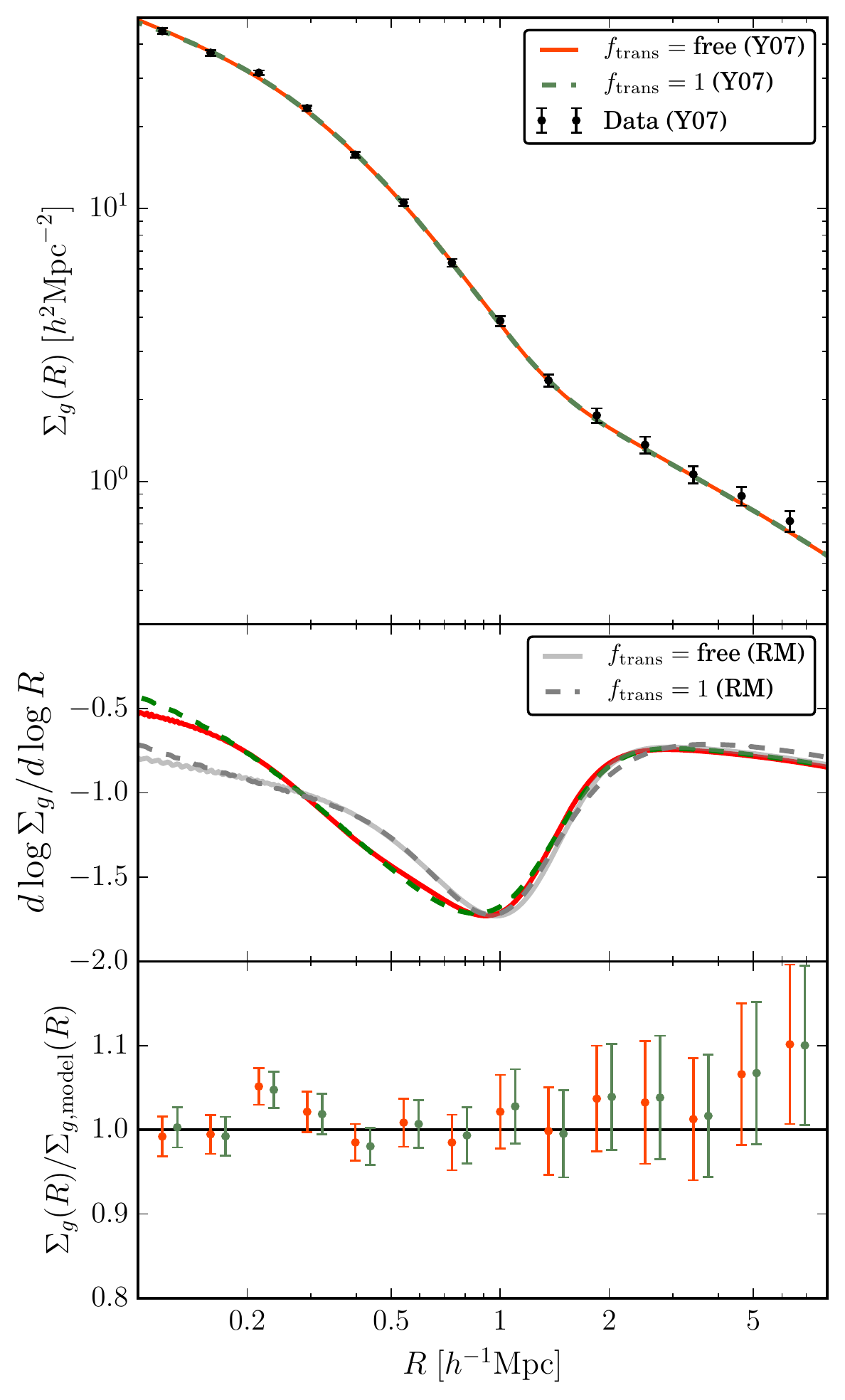}
\caption{Measurement of the splashback feature using the
  \citetalias{Yang2007} group catalog. The top panel shows the
  projected galaxy density profile $\Sigma_g$ overlaid with models
  with $f_{\rm trans}$ free (red solid) and $f_{\rm trans}=1$ (green
  dashed). The middle panel shows the log-derivative of
  $\Sigma_g$. We note that the two model fits are nearly identical
  in both panels. We also overlay in grey the same measurements shown
  in \Fref{fig:model_components_full}, which is based on the
  \redmapper (RM) cluster catalog. The feature around 1 $h^{-1} {\rm
    Mpc}$ in the \redmapper measurements appear slightly sharper than
  the \citetalias{Yang2007} group measurement. The bottom panel shows
  the ratio of the \citetalias{Yang2007} measurements to the best-fit
  models.}
\label{fig:model_components_yang}
\end{figure}

\begin{figure}
\centering
\includegraphics[width=0.85\linewidth]{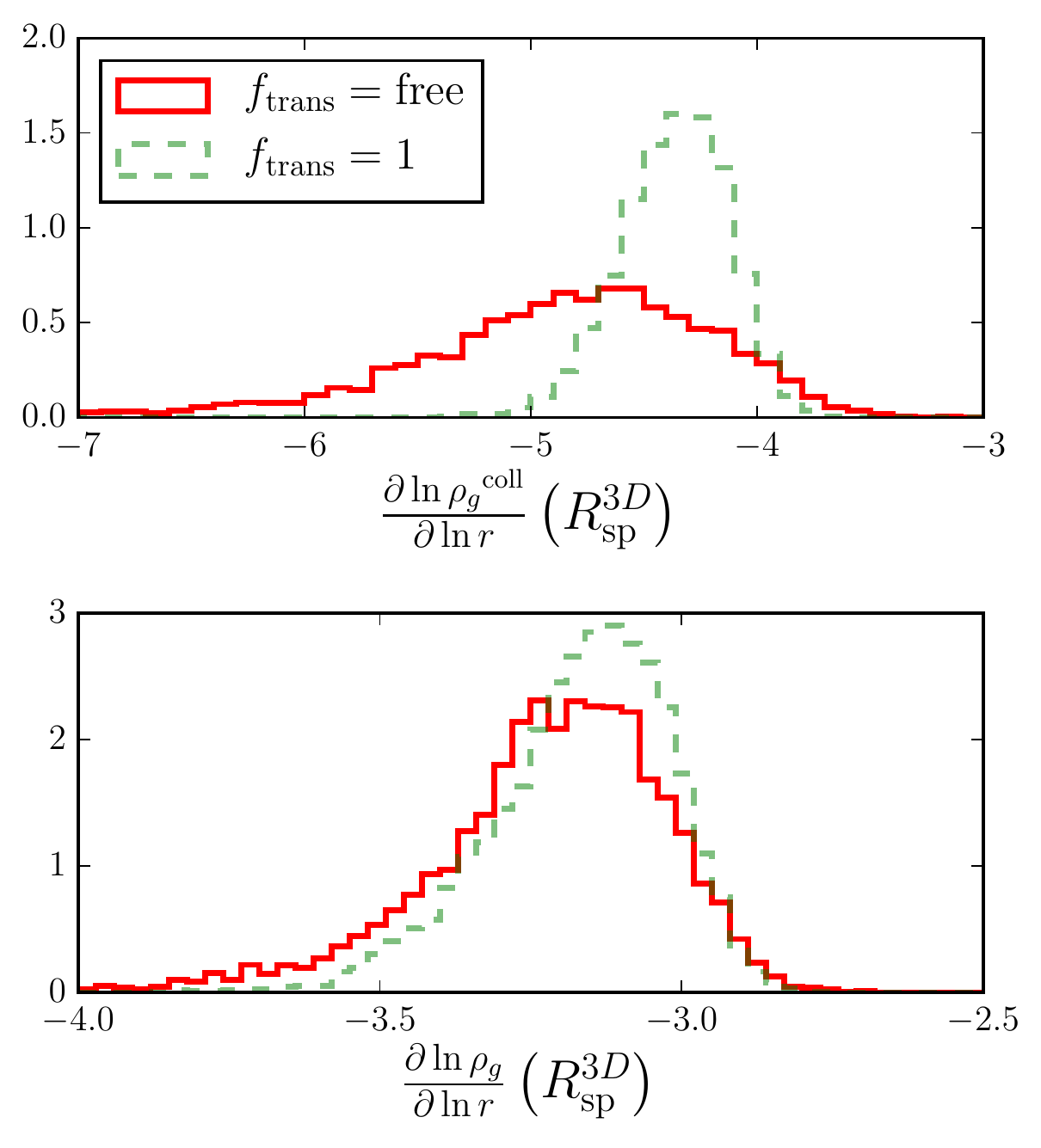}
\caption{Same as \Fref{fig:slope_dist}, but for the measurement of
  galaxy density around the groups in the \citetalias{Yang2007}
  catalog.}
\label{fig:slope_dist_yang}
\end{figure}

The measurement of galaxy density around the \citetalias{Yang2007}
groups is shown in \Fref{fig:model_components_yang}. The distribution
of logarithmic slopes at $R_{\rm sp}$ corresponding to this measurement
is shown in \Fref{fig:slope_dist_yang}.

\end{document}